\documentclass[12pt]{iopart}

\expandafter\let\csname equation*\endcsname\relax
\expandafter\let\csname endequation*\endcsname\relax

\usepackage{iopams}
\usepackage{latexsym}
\usepackage{amssymb}
\usepackage{amsmath}	
\usepackage{graphicx}
\usepackage{dcolumn}
\usepackage{bm}
\usepackage[usenames, dvipsnames]{color}
\usepackage{caption}
\usepackage{color}

\makeatletter
\def\blfootnote{\xdef\@thefnmark{}\@footnotetext}
\makeatother

\begin{document}

\title[Sr atom interferometry with the optical clock transition]{Sr atom interferometry with the optical clock transition as a gravimeter and a gravity gradiometer}

\author{Liang Hu$^*$, Enlong Wang, Leonardo Salvi, Jonathan N Tinsley, Guglielmo M Tino$^{\dagger}$ and Nicola Poli$^{\ddagger}$}
\address{Dipartimento di Fisica e Astronomia and LENS - INFN Sezione di Firenze\\
Universit\`{a} degli Studi di Firenze, Via Sansone 1, 50019 Sesto Fiorentino, Italy \\  }

\ead{nicola.poli@unifi.it}


\footnote[0]{$^*$ current address: Department of Electronic Engineering, Shanghai Jiao Tong University}
\footnote[0]{$^{\dagger}$ also at CNR-IFAC}
\footnote[0]{$^{\ddagger}$ also at CNR-INO}


\vspace{10pt}
\begin{indented}
\item[]
\end{indented}

\begin{abstract}
We characterize the performance of a gravimeter and a gravity gradiometer based on the $^{1}$S$_{0}$-$^3$P$_0$ clock transition of strontium atoms. We use this new quantum sensor to measure the gravitational acceleration with a relative sensitivity of $1.7\times10^{-5}$, representing the first realisation of an atomic interferometry gravimeter based on a single-photon transition. Various noise contributions to the gravimeter are measured and characterized, with the current primary limitation to sensitivity seen to be the intrinsic noise of the interferometry laser itself. In a gravity gradiometer configuration, a differential phase sensitivity of 1.53~rad/$\sqrt{\textnormal{Hz}}$ was achieved at an artificially introduced differential phase of $\pi/2$~rad. We experimentally investigated the effects of the contrast and visibility based on various parameters and achieve a total interferometry time of 30~ms, which is longer than previously reported for such interferometers. The characterization and determined limitations of the present apparatus employing $^{88}$Sr atoms provides a guidance for the future development of large-scale clock-transition gravimeters and gravity gradiometers with alkali-earth and alkali-earth-like atoms (e.g., $^{87}$Sr, Ca, Yb).
\end{abstract}

%
%
%
%
%

\section{Introduction}

Quantum sensors based on atom interferometry have attracted widespread research interest. Thanks to rapid development over recent decades, newly developed quantum sensors are expected to play a crucial role for science and technology in the near future~\cite{tino2014atom}. For example, atom interferometers based either on Raman or Bragg transitions have been widely employed in precision measurements of the Newtonian gravitational constant~\cite{rosi2014precision}, the local gravitational acceleration of Earth~\cite{peters1999measurement, hauth2013first, hu2013demonstration, gillot2014stability,asenbaum2017phase}, gravity gradients~\cite{snadden1998measurement, sorrentino2014sensitivity, d2016bragg} and gravity curvature~\cite{asenbaum2017phase,rosi2015measurement}, as well as in precision tests of the weak equivalence principle~\cite{fray2004atomic, tarallo2014test, zhou2015test, barrett2016dual, duan2016test, rosi2017quantum}.  Atom interferometers have also been considered for their possible application in the detection of gravitational waves~\cite{tino2007possible, dimopoulos2008atomic, yu2011gravitational}. In particular, several schemes based on the optical clock transitions of alkaline-earth-like atoms (e.g., Sr, Ca, Yb) have been recently proposed~\cite{graham2013new, kolkowitz2016gravitational, norcia2017role, canuel2017exploring} and in the case of Sr atoms have been tested in proof-of-principle experiments~\cite{hu2017atom, Katori2017}.

Single-photon atom interferometers could also be used to test the interplay between quantum mechanics and general relativity~\cite{rosi2017quantum,zych2011quantum, margalit2015self}, to test the weak equivalence principle with quantum superpositions of states with large energy ($\sim$eV) separation~\cite{rosi2017quantum}, and could enable low-energy table-top experimental tools for searches of new physics beyond the Standard Model~\cite{van2015search, arvanitaki2015searching, hamilton2015atom, safronova2017search,arvanitaki2018dm}.

In this article, we extend the study of an atom interferometer based on the $^{1}$S$_{0}$-$^{3}$P$_{0}$ optical clock transition of $^{88}$Sr atoms~\cite{hu2017atom}, demonstrating its application as a gravimeter and as a gravity gradiometer. We investigate the effect of laser phase noise on the interferometer, characterize its effect via the sensitivity function method and introduce a fibre-noise cancellation scheme. We experimentally study the interferometer contrast and fringe visibility, up to a total interferometry time of 30~ms, investigating the limitations imposed by several sources, such as the finite interferometry beam size, the atom number and the vertical velocity distribution of the atoms. We demonstrate that the amplified spontaneous emission coming from the laser system employed in the experiment contributes with a negligible effect on the interferometer, and that its performance can be improved by optimizing the interferometry laser beam size, atom number and vertical velocity distribution. Preliminary tests have also been performed on the clock transition of the fermionic $^{87}$Sr isotope, which has potential long-term advantages in comparison to $^{88}$Sr.




The paper is organized as follows: Section~2 introduces the main motivation and background for the new quantum sensor operated on the optical clock transition; Section~3 describes the experimental apparatus and procedures for both the gravimeter and the gravity gradiometer configurations; in Section~4 the main experimental results are presented, including a first assessment of the gravimeter sensitivity, a characterisation of the relevant parameters of the interferometry laser, the effect of the spatial properties of the atoms on the interferometer contrast and visibility, a further discussion of the gravity gradiometer configuration and its experimental realisation, and some preliminary measurements using $^{87}$Sr; finally, in Section~5 conclusions are given.

\section{Background}

\subsection{Motivation}

Most previous light-pulse atom interferometers have relied on multi-photon transitions to generate the superposition of momentum states required for sensitivity to inertial forces. These transitions, driven by Raman or Bragg beams, involve stimulated emission and absorption processes from two counter-propagating laser beams, which imprint their relative phase difference on the interacting atoms~\cite{tino2014atom}. Due to the finite speed of light, however, atoms at different spatial positions will interact with light emitted from the interferometric lasers at different times, meaning that any laser phase fluctuations on this timescale will be transferred to the interferometer and will not cancel even in a gradiometeric scheme~\cite{dimopoulos2008atomic,graham2013new}. For laboratory-scale experiments the resultant phase error is negligibly small, but for proposed long-baseline experiments, this noise term could begin to dominate~\cite{yu2011gravitational,graham2013new}. One potential application which would be affected by this noise is gravitational wave detection, where in order to attain the necessary instrument sensitivity, long baselines are usually employed, as for example in the LISA detector which is designed for a 2.5~$\times$10$^6$~km arm length~\cite{lisaWhitePaper_2017}.

Nevertheless, atom interferometers based on multi-photon transitions have been proposed as a means of detecting gravitational waves, both in space-based and in ground-based applications~\cite{yu2011gravitational, dimopoulos2009gravitational,hohensee2011sources, chaibi2016low}, although they will be severely affected by this noise at large scales. In this paper, in contrast, we study a novel atom interferometer based on the single-photon clock transition of atomic strontium ($^{1}$S$_{0}$-$^{3}$P$_{0}$), where this problem is absent. As the phase of the interacting photon is set at the point of emission from the laser, and doesn't acquire noise in the vacuum path between the two sensors, the common laser phase noise does not appear at the output of a single-photon gradiometer~\cite{yu2011gravitational, graham2013new}.

Despite this favourable feature, many technical challenges arise that motivate the study of the performance of interferometers operated on single-photon transitions. Indeed, because in the interferometer the atoms must spend a significant amount of time in the excited state, a narrow optical transition is required. Even though the requirements on the fractional frequency stability of the laser are relaxed in comparison to the multi-photon interferometer case, an ultra-stable laser is nevertheless needed to address such a transition. Moreover, because the atoms are in free flight, a high optical power is required to address atomic clouds with a finite velocity distribution width. Consequentially, the efficiency of the interferometer pulses will be studied both theoretically and experimentally throughout this paper.

\subsection{Rabi frequency}\label{subsec:rabiFreq}

The Rabi frequency is an important parameter that determines the efficiency of the interferometer pulse. Indeed, it is very important to operate with a high Rabi frequency, as for a given temperature of the atomic sample, the higher the Rabi frequency is, the greater the excited fraction of atoms will be.

For bosonic $^{88}$Sr atoms, the clock transition, which is otherwise forbidden, can be induced by a static magnetic field $B$ together with an optical field of intensity $I$. The corresponding Rabi frequency is given by~\cite{taichenachev2006magnetic},
\begin{equation}
\Omega_{88} = \alpha_{\text{Sr}}\sqrt{I}|B|,
\label{eq1}
\end{equation}
where $\alpha_{\text{Sr}}=2\pi\times198$ $\text{Hz}/[\text{T}\sqrt{\text{mW/cm}^{2}}]$. 

In our typical experimental conditions with a magnetic field $B=350$~G and a beam intensity $I=25$~W/cm$^{2}$, the achievable Rabi frequency is $\Omega_{88}\approx 2\pi\times1$~kHz.
This Rabi frequency is much smaller than the typical value obtained in Raman or Bragg interferometers ($\Omega_{R/B} > 2\pi\times100$~kHz). As the Rabi frequency scales linearly with the magnetic field intensity and with the square root of the laser intensity (Eq.~\ref{eq1}), an increase of the Rabi frequency by a factor of 10 is not easily attainable experimentally. For example, a beam intensity $I\approx2.5$~kW/cm$^{2}$ with a magnetic field $B\approx350$~G or a homogenous magnetic field of $B\approx 3500$~G with a beam intensity $I\approx25$~W/cm$^{2}$ would be necessary to obtain a Rabi frequency $\Omega_{88} \approx 2\pi\times10$~kHz. Unfortunately, all these requirements are experimentally challenging, particularly in the case of intense magnetic fields.  Moreover, a relative fluctuation of the magnetic field of only 0.1~\% at $3500$~G will induce a second-order Zeeman shift of about 8~kHz.  Similarly, a relative fluctuation on the clock beam intensity of only 0.01~\% at a peak intensity of 2.5~kW/cm$^{2}$ will lead to an AC Stark shift of 4.5~kHz. These large resonance frequency fluctuations, being comparable with the achievable Rabi frequency, can be difficult to compensate for and would induce phase shifts and instabilities on the excited fraction of atoms.

An alternative promising solution is to use the fermionic $^{87}$Sr isotope where the Rabi frequency is given by,
\begin{equation}
\Omega_{87} = \Gamma_{87}\sqrt{{I}/{2I_{s}}}
\label{eq.rabi87}
\end{equation}
where $I_{s}\approx 0.4$~pW/cm$^{2}$ and $\Gamma_{87}\approx2\pi\times 1$~mHz are the saturation intensity and linewidth of the $^{87}$Sr optical clock transition, respectively. In this case, to achieve a target Rabi frequency of $\Omega_{87}\approx2\pi\times10$~kHz, a clock beam intensity of $I=41$~W/cm$^{2}$ is sufficient and no intense magnetic field is required. Concerning the value of the Rabi frequency, the clock transition of fermionic Yb atoms (natural linewidth $\approx$~8~mHz) could also represent a valid alternative. 


While $^{87}$Sr atoms may seem a more promising solution in the long term, most of the work presented here is performed with the more abundant isotope $^{88}$Sr, with which a higher signal-to-noise ratio at detection has been experimentally obtained. This bosonic isotope also possesses other advantages due to the absence of nuclear spin, which results in efficient Doppler laser cooling down to the recoil temperature limit with a simple laser configuration and lower sensitivity of the clock transition to stray electromagnetic fields~\cite{boyd2007nuclear}. However, preliminary measurements of $^{87}$Sr Rabi oscillations have also been carried out showing the potential of the fermionic isotope (Section~\ref{subsec:sr87}).


\subsection{Gravimeter sensitivity}
A gravimeter based on a single-photon transition is sensitive to the phase of the photons which interact with the atoms. This is in contrast with the multi-photon case, in which the imprinted phase does not arise directly from the laser, but rather from the phase difference between the two counter-propagating beams, which are phase locked. In this configuration and for laboratory-scale interferometers, the laser phase noise is approximately common to the two beams and is therefore suppressed. Conversely, in a gravimeter utilising an optical clock transition, the laser must exhibit an extremely low phase noise, as it operates as a phase reference for the interferometer.



The accumulated phase difference $\Phi$ in a Mach-Zehnder single-photon gravimeter of total duration $2T+4\tau_{R}$ can be expressed as~\cite{antoine2006matter, li2015raman},
\begin{equation}
\Phi = (\frac{\omega_a}{c}g-\alpha)T^{2}\left[1+\left(2+\frac{4}{\pi}\right)\frac{\tau_R}{T}+\frac{8}{\pi}\left(\frac{\tau_R}{T}\right)^{2}\right]
+(\phi_{1}-2\phi_{2}+\phi_{3}) ,
\label{eq:phaseShift}
\end{equation}
where $\omega_a$ is the atomic transition angular frequency, $\tau_{R}$ is the duration of the $\pi/2$~pulse,  $\alpha$ is the frequency chirping rate applied to the interferometry laser for compensating the Doppler shift of the freely-falling atoms, and $\phi_{i}$ is the laser phase associated with the $i$-th interferometer pulse. 
As clearly shown by eq.~\ref{eq:phaseShift}, laser phase fluctuations will directly impact the phase difference accumulated in the interferometer, and will therefore be indistinguishable from changes in the gravitational acceleration. In repeated measurements, these instabilities will cause a loss of fringe visibility, which represents a major obstacle for precision devices, as is also the case for gravimeters based on multi-photon atom interferometry~\cite{hu2017atom, Katori2017, kasevich1991atomic,riehle1991optical}.

  \begin{figure}
    \centering
    \includegraphics[width=0.7\linewidth]{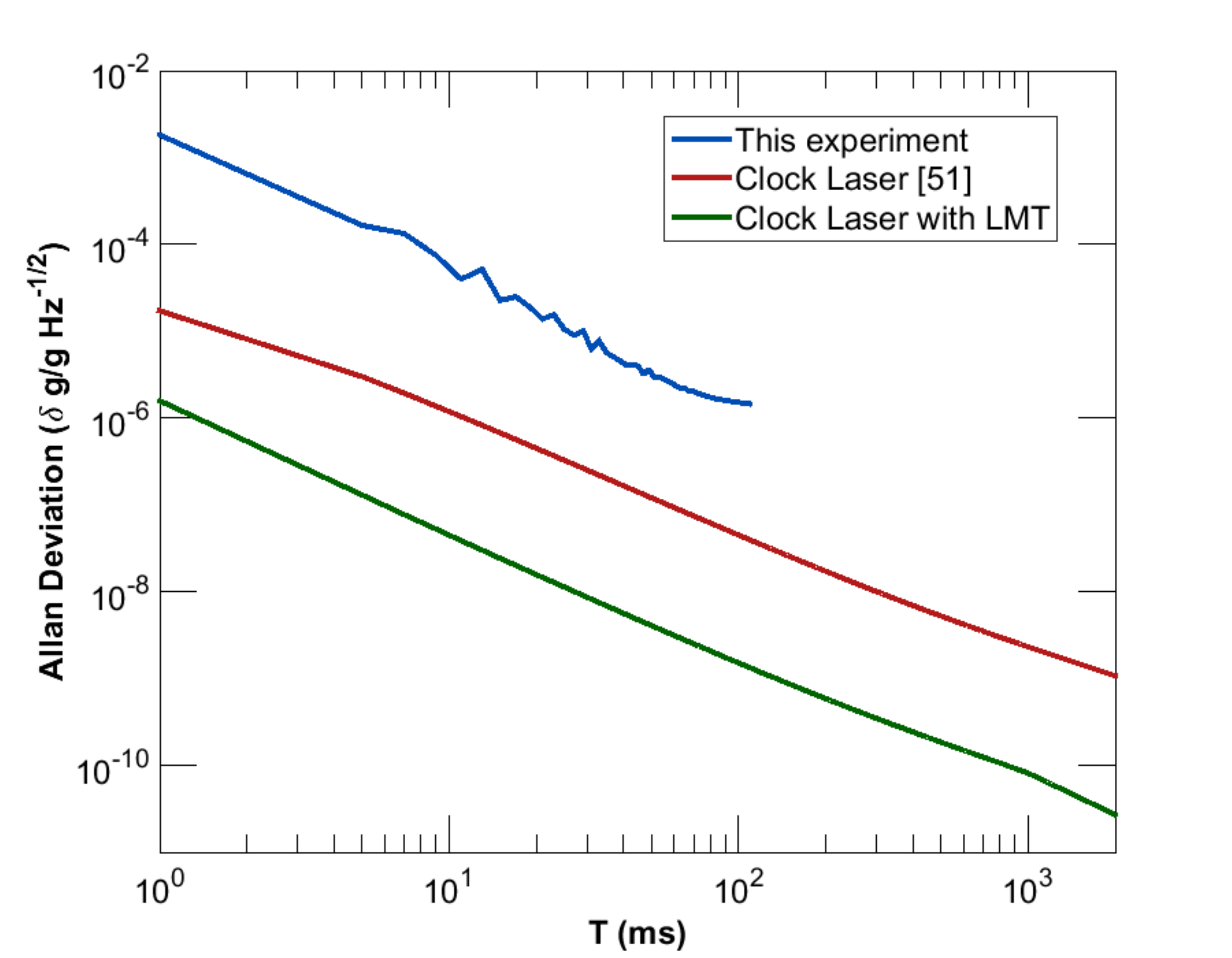}
    \caption{Calculated gravimeter sensitivity $\delta g/g$ at 1~s as a function of the interferometer laser phase noise. The calculation is made using the measured phase noise of the laser utilised in the experiment (blue line) and using the experimentally utilised interferometer parameters ($\tau$=450~$\mu$s, $T_c$=2T+2.4~s, $N$=1). This line is truncated at the time that the total phase noise equals $\pi/2$. Also shown are calculations based upon an estimate of the phase noise for a current state-of-the-art laser system, using both the experimental parameters (red line) and also for a future upgraded interferometer ($\tau$=100~$\mu$s, $T_c$=2T+1~s, $N$=20; green line).}
     \label{fig:gravSensitivityCalc}
\end{figure}

In general, the laser phase noise experienced by the atoms can be separated into two independent contributions $\delta\phi_A=\delta\phi_L+\delta\phi_{P}$, where $\delta\phi_{L}$ is the phase noise of the interferometry laser system, and $\delta\phi_{P}$ is the phase noise imposed by optical path fluctuations in the path delivering the laser to the atoms, for example by vibrations. In the case of Raman and Bragg atom interferometers, the phase reference is usually given by the retro-reflecting mirror used to create the counter-propagating beam configuration. By simply stabilising this retro-reflecting mirror, the deleterious effects of vibrations can be largely suppressed~\cite{zhou2012performance, hensley1999active,tang2014programmable}. In a single-photon atom interferometer, however, any optical component along the path connecting the clock laser to the atoms is a source of phase noise and the perfect isolation of all such optics is challenging. Nevertheless, schemes exist by which this noise ($\delta\phi_{P}$) can be largely removed, for example by introducing a secondary independent sensor to act as a phase noise detector~\cite{falke2012delivering,d2018measuring}, and we ignore it in the following calculations.

The effect of laser phase noise on the performance of a single-photon atom interferometer gravimeter can be quantitatively estimated through the sensitivity function method~\cite{cheinet2008measurement,le2008limits}. The Allan variance of the interferometric signal induced by the interferometry laser noise is given by~\cite{cheinet2008measurement,le2008limits},
\begin{equation}
	\sigma_{\phi}^{2}(\tau)=\frac{1}{\tau}\sum_{n=1}^{\infty}|H_\phi(2\pi n/T_c)|^2S_{\phi}(2\pi n/T_{c}),
	\label{eq:transferFunction}
\end{equation}
where $\tau$ is the averaging time, $H_{\phi}(f)=2\pi f\int_{-\infty}^{\infty}e^{i2\pi ft}g(t)dt$ with $g(t)$ being the sensitivity function for a Mach-Zehnder type interferometer~\cite{le2008limits}, and $S_{\phi}(f)=S_{\phi,O}(f)+S_{\phi, E}(f)$ is the total single-sided power spectral density (PSD) of the phase noise of the interferometry laser. The fractional sensitivity $\delta g/g$ is determined by measuring the phase fluctuations $\delta\Phi$ as a fraction of the total accumulated phase shift $\Phi$, as given by eq.~\ref{eq:phaseShift},
\begin{equation}
\frac{\delta g}{g}=\frac{\sigma_{\phi}(\tau)}{\Phi}.
\label{eq:deltaGoverG}
\end{equation}
 






To evaluate the potential performance of a single-photon atom interferometry gravimeter, we consider the achievable sensitivities based upon our setup and one using a leading laser system~\cite{matei201715, zhang2017ultrastable}, assuming that the sole contribution to the interferometer noise arises from the laser itself (Fig.~\ref{fig:gravSensitivityCalc}). These sensitivities are calculated for a variable interferometer time ($T$), but otherwise using both the experimental parameters utilised in the experiments presented in this article ($\tau$=450~$\mu$s, $T_c$=2T+2.4~s, $N$=1) and also for modest improvements ($\tau$=100~$\mu$s, $T_c$=2T+1~s, $N$=20), where $N$ is the relative amount of momentum imparted. In the last case, we simply scale the expected sensitivity by the value of the large-momentum transfer to achieve an estimate, though for a full calculation a multi-pulse implementation of the transfer function must be considered~\cite{Decamps_2018}, as such an enhancement requires a sequence of pulses to be sent~\cite{graham2013new}.

These calculations highlight the importance of utilising an ultra-stable laser for the gravimeter, as the laser used in the experiments presented in this paper reaches a phase noise of $\pi/2$ after around 150~ms of interferometer time, the level at which it is difficult to extract meaningful results due to imprecise knowledge of the position on the interference fringe. This is despite this laser having a linewidth of the order 1~Hz~\cite{tarallo2011high}. However, using state-of-the-art optical clock lasers, it should in principle be possible to reach sensitivities at the 10$^{-9}$ level at reasonable interferometry times. Such a laser operating with improved experimental values and utilising large-momentum transfer techniques, could also in principle reach a performance comparable to the current leading multi-photon gravimeter, which operates at a sensitivity of $3.0 \times 10^{-11} /\sqrt{\text{Hz}}$, at T~=~1.15~s~\cite{dickerson2013multiaxis}. We note, however, that such experiments are not usually limited by phase noise, but other factors such as detection performance or atom shot noise.

\subsection{Gravity gradiometer}
When two vertically separated atom interferometers are operated simultaneously with the same laser beam on the optical clock transition, a gravity gradiometer can be implemented. Contrary to the case of the gravimeter, this configuration should be highly insensitive to the laser’s intrinsic phase noise. Such a cancellation is normally also expected for interferometers based on multi-photon Raman or Bragg transitions. In this case, however, and as mentioned earlier, if the separation between the two interferometers is large, then in order to guarantee that the light from two counter-propagating lasers reaches the two clouds at the same time, a time delay between the times when the laser pulses are sent from the two sides must be introduced. For a cloud separation $L$, this delay amounts to $\Delta t = L/c$ and laser phase fluctuations occurring during this time no longer cancel and instead enter as a differential effect. For laboratory-scale experiments, where the cloud separation is on the order of 10~cm, this results in the laser’s phase noise above a frequency $1/\Delta t$ of 3~GHz entering the interferometer output. For a typical laser linewidth of less than 1~MHz and typical laser platform vibrations at frequencies below 10~kHz, this noise contribution can usually be neglected. 

However, when a large-scale gradiometer is considered, the delay $\Delta t$ can introduce a dominant noise contribution. For example, in order to enhance the sensitivity to the point where gravitational waves are detectable, gradiometer baselines on the order of 100,000~km may be considered. In this case, laser noise above only 3~Hz would contribute to the gradiometric signal. This would mean that the laser’s intrinsic noise should be reduced by locking to state-of-the-art ultrastable cavities and that the laser platform motion should be limited by vibration isolation. In particular, to implement a gravitational wave detector with atoms reaching a strain sensitivity of $h=10^{-20}$, a laser frequency stability on the order of $\delta \omega/\omega \approx h$ is required for first diffraction order. This stability is far from the present state-of-the-art in laser frequency stability~\cite{matei201715, zhang2017ultrastable}.
 
On the other hand, the gradiometer operating on the optical clock transition should be insensitive to both the laser's intrinsic phase fluctuations and the platform vibrations. This can be intuitively understood by recognizing that no temporal superposition is required when a transition is driven with a single laser and that a single photon does not acquire extra phase noise in the (vacuum) path between the two interferometer clouds.
In our proof-of-principle experiment, a gradiometer was implemented with a relatively small size. Specifically, a cloud separation of $\Delta r = 1.9$~mm with an interferometer time $2T = 10$~ms was obtained~\cite{hu2017atom}. The predicted differential phase shift induced by Earth's gravity gradient $\Gamma = 3.1\times 10^{-9}$~s$^{-2}$, for negligible velocity difference between the two clouds, i.e. $\Delta\Phi \approx \frac{\omega_a}{c}\Delta r\Gamma T$, is only on the order of 1~nrad, well below our current sensitivity level. As a result, in order to characterize the sensitivity of our gradiometer, a known artificial differential phase shift was imprinted onto the atoms. This was attained by introducing a velocity difference between the interferometer clouds and by addressing them with two different frequency components of the interferometer laser. By introducing a known phase shift between these two components, the imprinted phase shift could be tuned at will, allowing a measurable signal to be obtained and characterized.

\section{Experimental apparatus and procedure}

\subsection{Experimental apparatus}
To perform atom interferometry with the Sr clock transition several important experimental requirements must be met. A large number of Sr atoms must be trapped and cooled towards ultra-cold temperatures, before interacting with a laser with sufficiently narrow linewidth and optical power to drive the clock transition. Maximum sensitivity of the interferometer is achieved by increasing the total interferometry time, realised experimentally by having the capability of launching the atom cloud upwards. Finally, the apparatus must be capable of detecting and resolving the two output states of the interferometer, namely the $^{1}$S$_{0}$ and $^3$P$_0$ states of Sr. To maximise the signal-to-noise ratio, only those atoms which have undergone the interferometry sequence should be detected, meaning the capability to remove unwanted atoms is needed.

\begin{figure}
    \centering
    \includegraphics[width=0.75\linewidth]{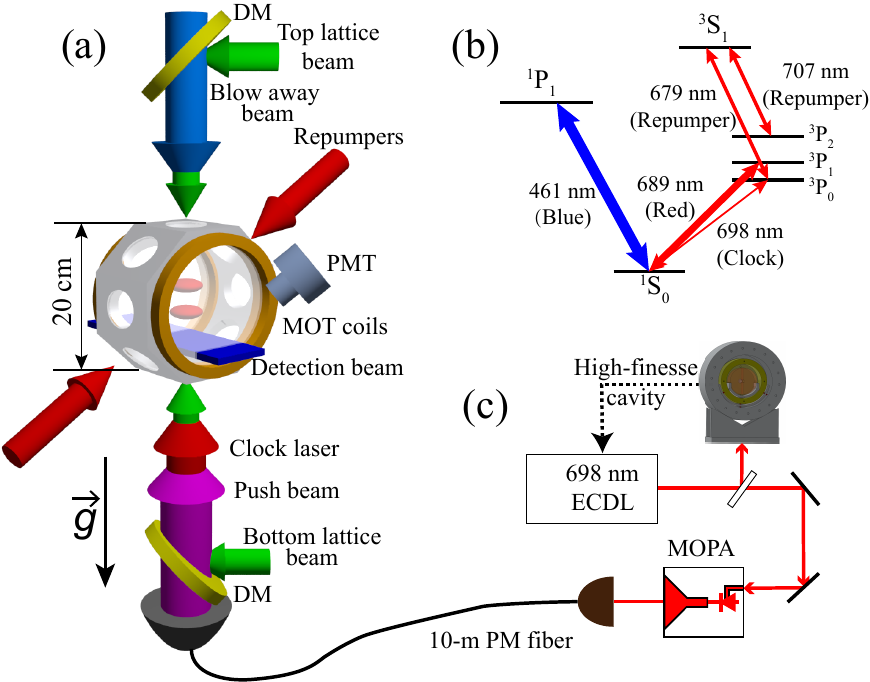}
    \caption{  (a) Experimental apparatus. Ultra-cold strontium atoms are produced in a magneto optical trap (MOT) inside a 20-cm-diameter science chamber. Various laser sources are required: a (clock) interferometry beam; two 1D green lattice beams; a blue detection beam; a red push beam; a blue blow-away beam; and two red repumper beams. A homogeneous static magnetic field produced by the MOT coils is used to induce the clock transition of $^{88}$Sr atoms.  (b) Relevant strontium atomic energy levels and transitions, including cooling transitions at 461~nm (blue) and 689~nm (red), repumper transitions at 679~nm and 707~nm and clock transition at 698~nm (clock). (c) Simplified view of the interferometry laser setup. Laser radiation at 698~nm is frequency stabilized in two steps via Pound-Drever-Hall (PDH) locking to optical cavities. The power of the stabilized light is then boosted by a semiconductor master oscillator power amplified (MOPA) laser setup and sent to the science chamber through a 10~m long polarization-maintaining (PM) fiber. Other abbreviations: ECDL, external-cavity diode laser; DM, dichroic mirror; PMT, photomultiplier tube.}
     \label{fig1}
\end{figure}

A schematic view of the experimental apparatus utilised to achieve these requirements is shown in Fig.~\ref{fig1}~(a). It mainly consists of a science chamber, where the atoms are cooled and trapped in a magneto-optical-trap (MOT); a green laser which is used to produce an accelerating standing wave trap; a semiconductor master oscillator power amplified (MOPA) interferometry laser, used to induce clock transitions between the two interferometer states; a blue blow-away beam; a detection setup including a detection beam and a photomultiplier tube (PMT); a red push beam; and two red repumper beams, which keep the atoms from falling into dark states and are also used to have ensure all the atoms are in the ground state for detection.

The core apparatus for the MOT preparation stage has been described previously~\cite{poli2014transportable, zhang2016trapped}. In brief, a cloud of $\approx 5\times10^{6}$ ultra-cold $^{88}$Sr atoms at a temperature of $\approx$1~$\mu$K, with a full-width at half-maximum (FWHM) in the horizontal (vertical) dimension of 300~$\mu$m (70~$\mu$m) is produced by a two-stage magneto-optical trap (MOT), with the cold atoms provided by means of a Zeeman slower.

The master laser at 698~nm is an external cavity diode laser (ECDL) (Sacher, SAL-0690-025) that is frequency stabilized in two steps with medium and high finesse ($\mathcal{F}=10000$ and  $500000$, respectively) Fabry-P\'{e}rot cavities. A beat note with a FWHM of 1~Hz has been previously observed for this system, when comparing two independent systems~\cite{tarallo2011high, poli2014transportable}. To provide a sufficient power level on the clock transition, the output power of the stabilized laser is further boosted by a MOPA laser consisting of a slave laser and a tapered amplifier (Eagleyard, EYP-TPA-0690). The output from the tapered amplifier is coupled into a 10-m polarization-maintaining (PM) single-mode optical fiber. At the fiber output, the beam is magnified and collimated by a set of telescopes and aligned into the atomic fountain along the vertical direction. The application of this beam to the atoms is controlled by an acousto-optical modulator (AOM) and the maximum power that can be delivered to the atoms is 80~mW.
 
 
Along the same direction, a vertical 1D lattice is used to hold and launch the atoms upwards, as previously described in Ref.~\cite{zhang2016trapped}. In brief, however, the output of the laser source at 532~nm (Coherent Verdi-V6) is equally split into two beams, with each beam passing through AOMs to independently control their frequency difference and relative amplitude. Each lattice beam is then coupled into independent PM fibers and sent from opposite directions to the atoms with the same linear polarization. The two lattice beams are superimposed onto the path of the interferometry laser by means of dichroic mirrors. With a power in each beam $P_g\approx1$~W and a $1/e^{2}$ beam radius $w_g\approx$ 350~$\mu$m, corresponding to a Rayleigh length $z_g\approx72$~cm, the trap depth at the position of the atoms is $U_0\approx9E_{r,g}$ (where $E_{r,g}=\hbar^{2}k_{g}^{2}/2m\approx 0.8\,\mu$K is the recoil energy in temperature units of the green lattice, $k_{g}$ is the wave number of the lattice light and $m$ is the mass of $^{88}$Sr atoms). The verticality of the interferometry laser and of the green lattice beams has been verified to within a few mrad by utilising the reflection from a water surface~\cite{tarallo2014test, mazzoni2015large, zhang2016trapped}.

The blue blow-away beam resonant with the strong $^{1}$S$_{0}$-$^{1}$P$_{1}$ transition at 461~nm is also superimposed onto the same vertical direction. This beam has an intensity of 3~mW/cm$^{2}$ and is used to remove unwanted atoms from those which have been selected for the interferometry sequence, improving the signal-to-noise ratio at detection.

The detection beam is tuned to the $^{1}$S$_{0}$-$^{1}$P$_{1}$ transition at 461~nm with a power of $\approx$~500~$\mu$W. This beam shares the same windows with the horizontal MOT beams, it is located 1.4~cm below the center of the MOT beams and is retro-reflected to enhance the signal size. This configuration allows for a maximum time-of-flight (TOF) of $\approx$~40~ms. The detection beam size is about 5~mm and 200~$\mu$m diameter in the horizontal and vertical direction, respectively. This aspect ratio guarantees a uniform interaction with the expanded cloud along the horizontal direction and the maximum vertical resolution. The fluorescence from the atoms is collected by a photomultiplier module (PMT).  

In the current setup, due to the typically small time of flight of the atoms from the end of the interferometer sequence to the final detection, together with the small momentum difference ($\hbar k$) between the two interferometer states, atoms in the two states are insufficiently spatially separated to allow for resolvable state detection. In order to mitigate this problem, we shine a push beam, resonant with the $^{1}$S$_{0}$-${^{3}}$P$_{1}$ transition at 689~nm and with a power of $\approx$ 2.8~mW and a diameter of 2~mm, upwards from the bottom of the chamber. This beam decelerates the atoms in the ground state, spatially separating them from the atoms in the excited state. After the push beam pulse and right before final detection, atoms in the excited state are eventually repumped back to the ground state by turning on the two repumper beams at 679~nm and 707~nm for $\approx$~10~ms, with the beams derived from extended-cavity diode lasers. 

\subsection{Gravimeter procedure} \label{subsec:gravProceduce}
  \begin{figure}
    \centering
    \includegraphics[width=0.8\linewidth]{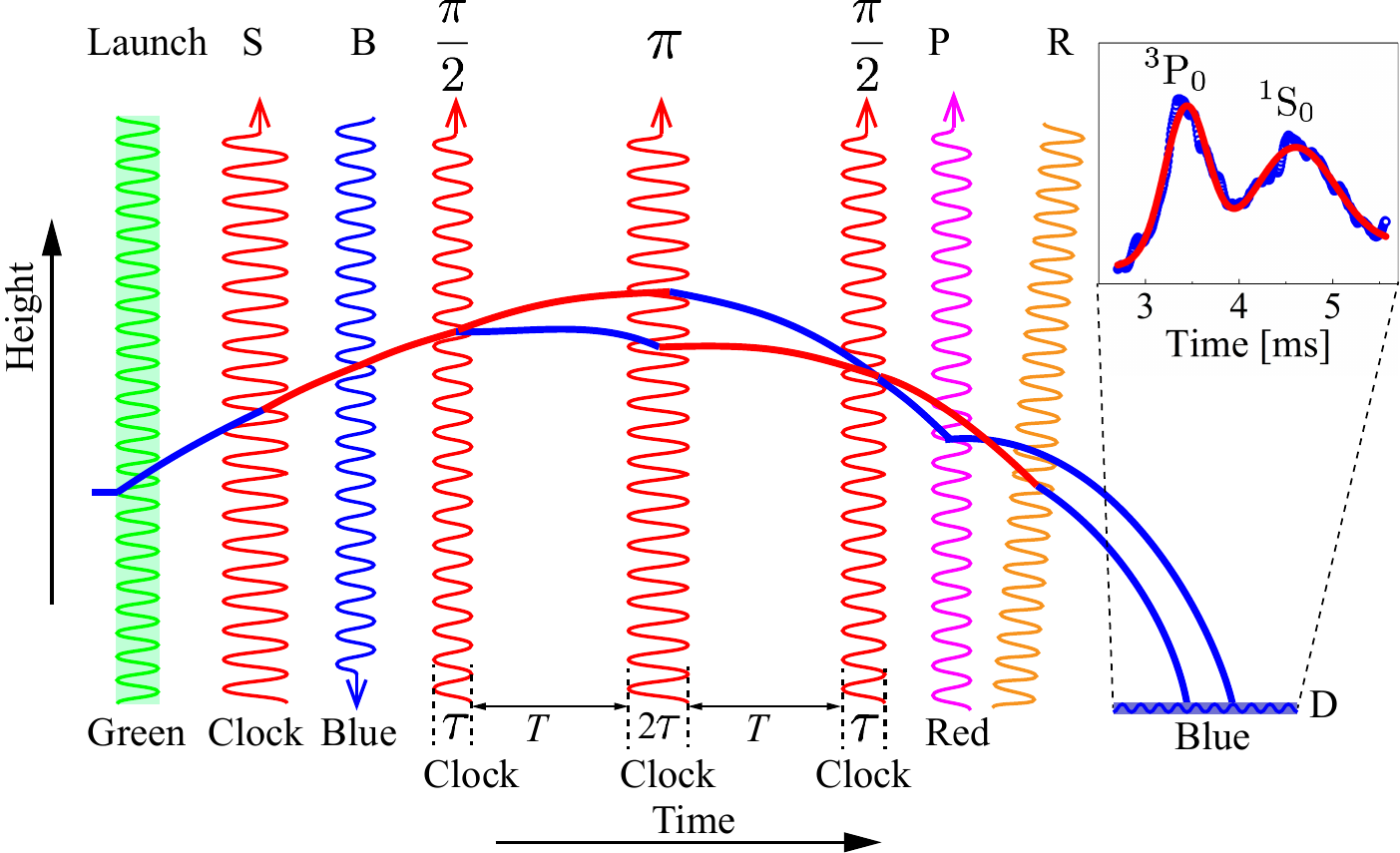}
    \caption{Single-photon gravimeter experimental sequence. Following loading into the lattice, the atoms are elevated upwards in a fountain up to an apogee of 0.8~cm above the MOT position. Next, a velocity selection $\pi$ pulse on the clock transition is applied (S), followed by a blow away beam (B) to leave a vertically cold sample. Subsequently, a standard Mach-Zehnder interferometer pulse sequence consisting of three interferometry laser pulses separated by two equal time intervals $T$ is performed.  At the end of the interferometric sequence, a push beam (P) is used to spatially separate the two interferometer states, followed by a repumper (R) pulse, in order to return all the atoms into the ground state for detection. Both arms of the gravimeter are finally detected by a horizontal detection beam (D) resonant with the strong dipole allowed transition of strontium. The inset shows a typical detection signal. A Gaussian fit of the signal peaks resolves the number of atoms in each state, giving the relative population for the gravimeter.}
     \label{fig:gravSeq}
\end{figure}

A typical experimental sequence of our single-photon gravimeter is shown in Fig.~\ref{fig:gravSeq}.  After the MOT preparation stage, about 10\% of the atoms are adiabatically loaded into the 1D lattice.  Whilst the atoms are being loaded into the lattice, the MOT beam intensity is reduced and the second-stage-MOT laser frequency is shifted closer to the unperturbed resonance (detuning $\Delta_r=-30$ kHz), in order to account for the light shift induced by the lattice light. This procedure reduces the losses due to the initial evaporation of hot atoms from the lattice, and results in an almost three-fold improvement in the number of atoms available for the interferometer~\cite{zhang2016trapped}.

Once the atoms have been loaded into the lattice, they are launched vertically upwards in order to increase the total interferometer time. The atoms are elevated upwards by chirping the frequency difference between the two lattice beams from 0 to 650~kHz with a rate of 10~kHz/ms, corresponding to a constant acceleration of 2.7~m/s$^{2}$. Under these conditions, 70\% of the atoms are elevated to a height of about 8~mm above the MOT position. During the lattice launch sequence, which lasts for 65~ms, we invert the current direction in one of the MOT coils to generate the homogeneous magnetic field ($|B|\approx350$~G) necessary for inducing the clock transition of $^{88}$Sr atoms~\cite{taichenachev2006magnetic}.  Subsequently, the remaining atoms are released from the trap and a $\pi$ pulse on the clock transition is applied to select a narrow-width velocity class~\cite{hu2017atom, moler1992theoretical}. The vertical velocity distribution and the atom number of the selected cloud are set by the duration of this selection pulse (see Section~\ref{subsec:velSel}). Atoms which are not selected by this $\pi$ pulse remain in the ground state and are removed by the application of the blue blow-away beam for 100~$\mu$s.

We then apply a standard Mach-Zehnder interferometer pulse sequence consisting of three interferometry laser pulses separated by two equal time intervals $T$ (i.e., a first $\pi/2$ pulse to split, a middle $\pi$ pulse to reflect, and a final $\pi/2$ pulse to close the interferometer). Once the interferometry sequence is completed, the final population of the ground and excited states are detected.

We deduce the interferometer phase shift by using the measured relative population between the two clock states (see inset of Fig.~\ref{fig:gravSeq}), defined as $N_{\text{rel}}=N_{^{1}\text{S}_{0}}/(N_{^{1}\text{S}_{0}}+N_{^{3}\text{P}_{0}})$, where $N_{^{1}\text{S}_{0}}$ and $N_{^{3}\text{P}_{0}}$ are the total atom number in the states $^{1}\text{S}_{0}$ and $^{3}\text{P}_{0}$, respectively. The relative population is related to the interferometric phase $\Phi$ by the standard formula,
\begin{equation}
N_{\text{rel}}=P_{0}+\frac{V}{2}\cos(\Phi)
\label{eqshot}
\end{equation}
where $P_{0}$ is an offset, $\Phi$ is defined in Eq.~\ref{eq:phaseShift} and $V$ is the visibility.


By scanning the phase of the final, recombination pulse $\phi_{3}$, oscillations in relative atom number $N_{\text{rel}}(\Phi)$ can be observed. Here the interferometer contrast is estimated by the dispersion of measured values $N_{\text{rel}}(\Phi)$ from the 2nd to the 98th percentile, while the fringe visibility $V$ is estimated by the amplitude of the fitted sinusoidal function on each data set~\cite{hu2017atom}.

\subsection{Gravity-gradiometer procedure} \label{subsec:gravGradProcedure}

 \begin{figure}
    \centering
    \includegraphics[width=0.8\linewidth]{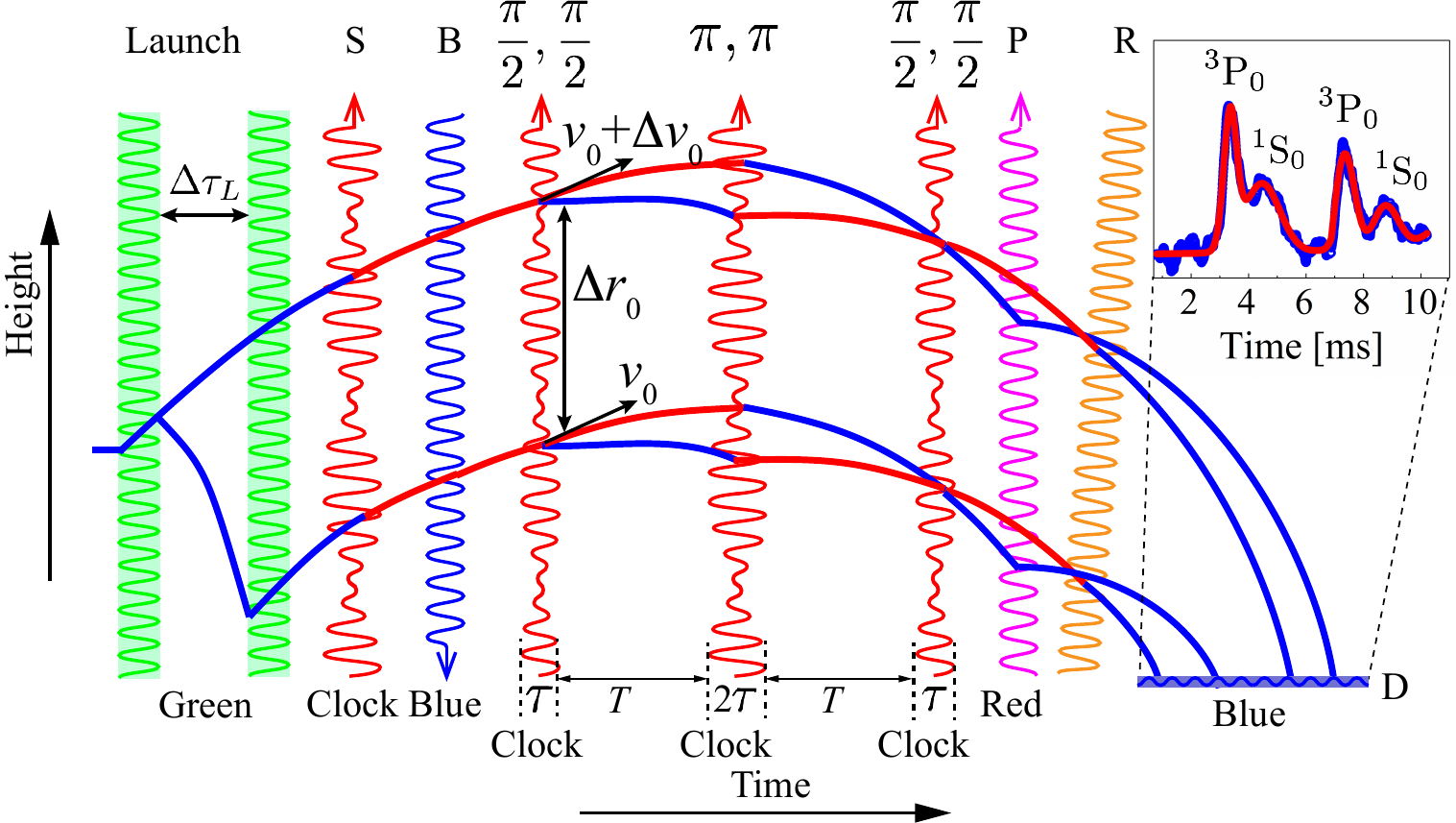}
    \caption{Single-photon gravity gradiometer experimental sequence. Two $^{88}$Sr atomic clouds are launched vertically in a fountain with a double launch technique from the same MOT cloud. A single $\pi$ selection pulse (S), but containing two different frequency components, is applied, resulting in two clouds with a vertical separation of $\Delta r_{0}=1.9$~mm and a relative vertical velocity difference of $\Delta v_0=6.5$~mm/s. Following this, the applied sequence of pulses is equivalent to the case of the gravimeter (Fig.~\ref{fig:gravSeq}), except that the interferometry pulses contain two frequency components. The inset shows a typical detection signal, with the two arms of each accelerometer resolved in time, for a total of four ports. A Gaussian fit of the signal peaks is used to estimate the number of atoms in each state, giving the relative population for each accelerometer.}
     \label{fig:gravGradSeq}
\end{figure}

The space-time trajectory of a gravity gradiometer is depicted in Figure~\ref{fig:gravGradSeq}. Here, slight modifications to the gravimeter experimental sequence and apparatus configuration outlined above are made~\cite{hu2017atom}. 
While the cloud preparation stage is similar to that of the gravimeter, in this case, after loading the atoms into the vertical green lattice and turning on the B field for inducing the clock transition, a double lattice launch technique~\cite{delaguila2018bragg} is adopted to produce two vertically separated atomic clouds.  The separation of the two clouds depends on the relative duration of the two launches and on the delay time $\Delta \tau_L$ between the two launches. For a duration of 8~ms and 6~ms for the first and second lattice launches, respectively, maintaining a chirp rate of 100~kHz/ms and with a delay time $\Delta\tau_L=$ 1~ms, the double lattice launch produces about $3.5\times10^{5}$ atoms in each cloud with a vertical separation of $\Delta r_{0}\approx$  1.9~mm (see Section~\ref{subsec:latticeLaunch}). It is clear that in this situation, the two clouds will also have a velocity difference $\Delta v_{0}\approx 6.5$~mm/s. Due to the different velocities, the resonance frequency differs by $\delta\omega=k\Delta v_0\approx 2\pi \times 9.3$~kHz between the two interferometers, which is significant compared to the Rabi frequency. 

To satisfy this condition, the interferometry laser spectrum contains two frequency components. This is obtained by inserting an AOM on the interferometry laser path which is driven by two radio frequency (RF) signals at frequencies of 80~MHz $\pm$ $\delta\omega/4\pi$ and with equal amplitude. The RF signals are produced by a two-channel direct digital synthesizer (DDS) generator 
and their relative phase shift $\delta\phi$ can be precisely tuned by an internal phase shifter. 

The two generated optical frequencies differing by $\delta\omega$ follow the same optical path; thus, phase fluctuations that may be caused, for example, by vibrations or air currents are common-mode, and do not degrade the performance of the gradiometer. The measured integrated phase variance between the two channels on the RF generator is about 1~$\mu\text{rad}^{2}$ and can therefore be neglected. The power in each component is set to half of the total interferometry laser power, in order to guarantee the same Rabi frequency for both accelerometers.

After the double launch, an experimental sequence similar to the gravimeter one, is adopted. In outline, a composite pulse sequence, consisting of an initial selection $\pi$ pulse, followed by a standard $\pi/2-\pi-\pi/2$ interferometer pulse sequence, is applied to both clouds. Each pulse includes two frequency components as indicated in Fig.~\ref{fig:gravGradSeq}. At the end of the gradiometer sequence, four output ports fall through the detection beam at different times, allowing the determination of the acquired phase for each interferometer. A typical detection signal for a gravity gradiometer is shown in the inset of Fig.~\ref{fig:gravGradSeq}.

\section{Experimental results}

\subsection{Interferometry laser characterization}

\subsubsection{Tapered amplifier characterization}
As discussed in Section~\ref{subsec:rabiFreq}, maintaining a high Rabi frequency is of crucial importance to the performance of the interferometer. For a constant beam diameter, the Rabi frequency is increased by increasing the optical power of the interferometry laser. Currently the maximum power available for the interferometry beam is limited by the attainable output power of the MOPA and the achievable fibre-coupling efficiency. Both of these parameters depend strongly upon the behaviour and performance of the tapered amplifier (TA), which is relatively uncommon and untested at 698~nm.

\begin{figure}[t]
    \centering
    \includegraphics[width=\linewidth]{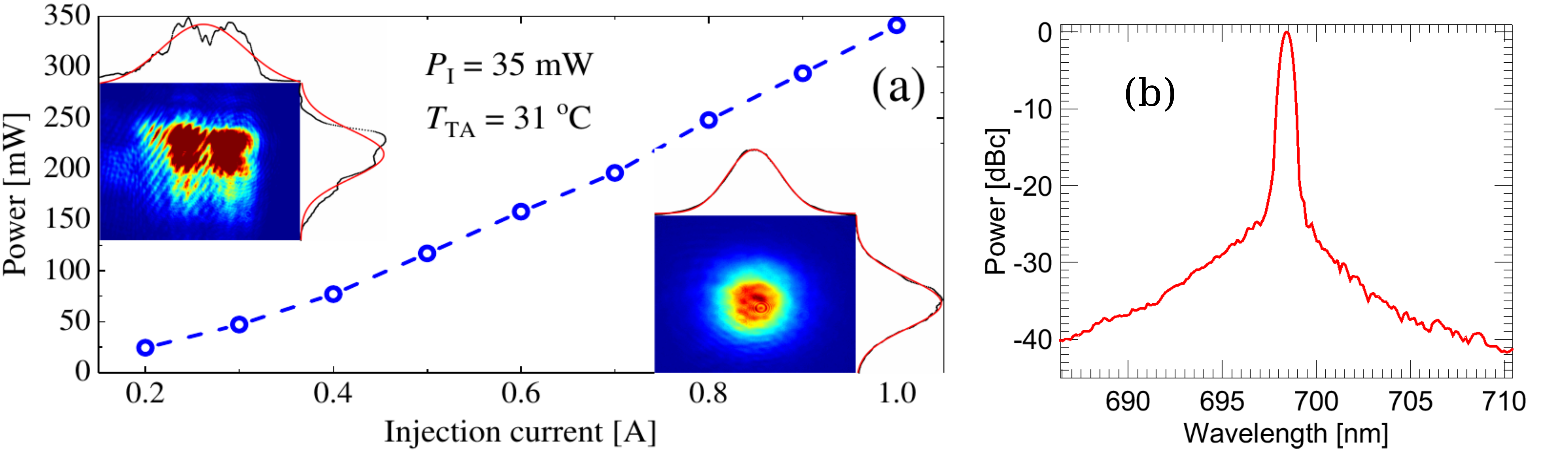}
    \caption{(a) Measured TA output power as a function of the TA operation current $I_{\text{TA}}$ at an injection power $P_{\text{I}}$ = 35 mW, and at a TA temperature $T_{\text{TA}}$ = 31 $^{\circ}$C. Left and right insets show the spatial intensity profiles of the clock laser beam at the TA output and after the 10-m PM fiber, respectively, with Gaussian fits to the parallel and perpendicular beam profile cross-sections also shown. (b) A typical emission spectrum of the clock laser after the  fiber measured by a home-made grating spectrometer with a resolution of $0.25$~nm. The broad ASE background is distinguished from the narrow peak of the signal.}
    \label{fig:taCharacterisation}
\end{figure}

We therefore performed a series of tests on the TA to characterize its gain and saturation level as a function of the TA operating temperature ($T_{\text{TA}}$), current ($I_{\text{TA}}$) and the optical injection power ($P_{\text{I}}$). As shown in Fig.~\ref{fig:taCharacterisation} (a), a maximum amplification of about 10~dB, up to a total output power of 340~mW, was reached at $P_{\text{I}}$ = 35~mW, $I_{\text{TA}}$ = 1~A and $T_{\text{TA}}$ = 31~$^{\circ}$C. 
Under these conditions, as shown in the left inset of Fig.~\ref{fig:taCharacterisation}~(a), the TA output has a non-Gaussian beam intensity profile with M$^2\approx$~2.4. In order to have a more homogeneous phase reference for the interferometer, the beam intensity profile is cleaned by the 10-m PM fiber and the output with M$^2\approx$~1.2 is shown in the right inset of Fig.~\ref{fig:taCharacterisation}~(a). While initially more than 60\% of power has been coupled, aging effects in the TA after 100~hours of work resulted in a dramatic reduction of the coupling efficiency down to 45\%.  The eventual result is that a maximum power of 80~mW can be delivered to the atoms, after passing through the 10-m PM fiber and a subsequent AOM for pulse shaping. One potential way to increase the available power of this system would be to combine several MOPAs by using a heterodyne optical phase-locked loop technique~\cite{liang2007coherent2}.

\subsubsection{Photon scattering}



  \begin{figure}
    \centering
    \includegraphics[width=1\linewidth]{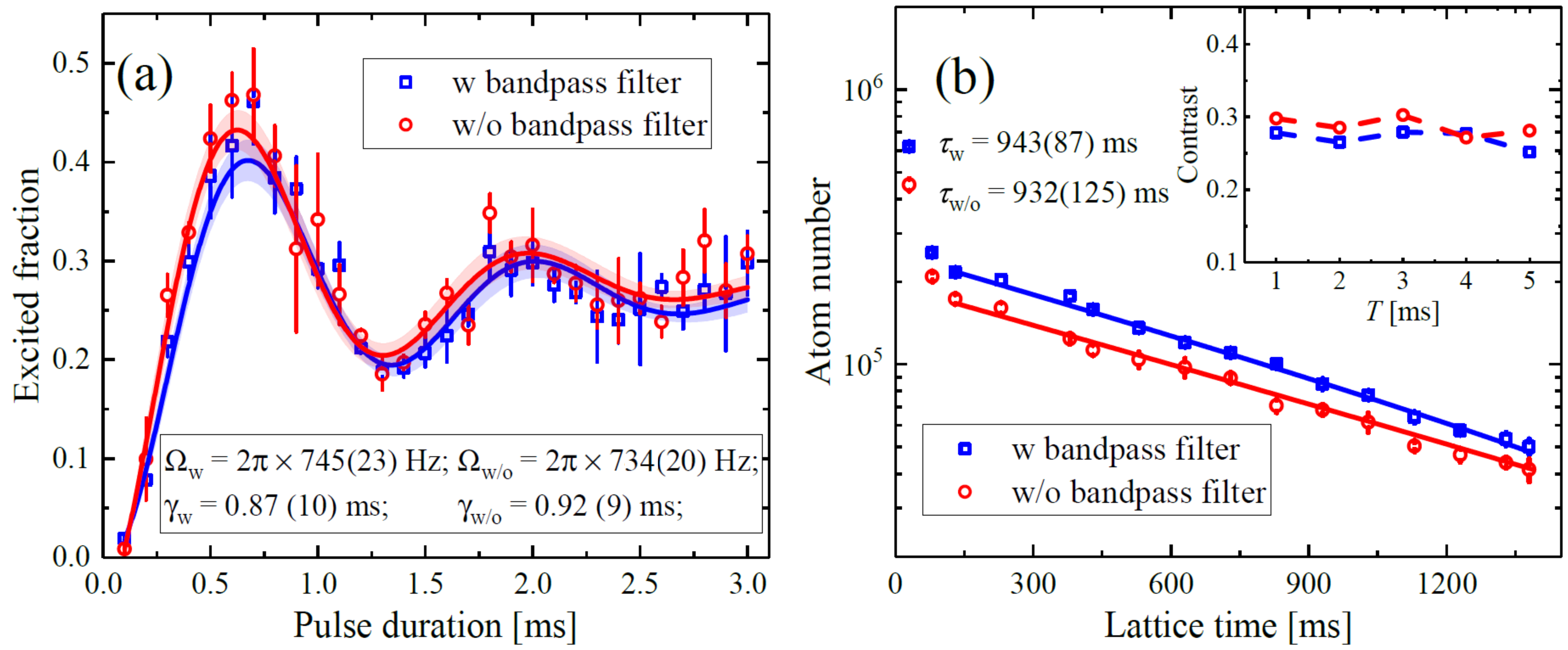}
    \caption{Excited fraction of atoms as a function of interferometry pulse duration observed with (blue squares) and without (red circles) a bandpass filter used to reduce the ASE radiation at 689~nm. The results are fitted by a damped sinusoidal function with the shadow indicating the 95\% confidence interval of the fit. The Rabi frequencies and damping times show no significant change between the two cases. (b) Measurement of the lifetime of atoms trapped in a vertical lattice at 532~nm in the presence of interferometer laser with (blue squares) and without (red circles) the bandpass filter, where a small offset in atom number has been introduced for clarity. The solid lines indicate exponential decay fits with estimated decay time constants of 943(87)~ms and 932(125)~ms, indicating a negligible effect from ASE radiation.  The inset shows the measured interferometer contrast as a function of interferometer time $T$ with (blue squares) and without (red circles) the bandpass filter.}
    \label{fig:bandpass}
\end{figure}


In Raman or Bragg interferometer pulses, spontaneous emission constitutes a major limitation to the interferometer contrast, and short pulse durations and large detunings from the single-photon transition are usually needed~\cite{kasevich1991atomic, mazzoni2015large, delaguila2018bragg}. Although the loss of contrast due to spontaneous emission is largely reduced when directly utilising single-photon transitions due to the virtually infinite lifetime of the upper state~\cite{Poli2013, Ludlow2015}, additional contributions due to resonant scattering from other excited states remain a potential source of decoherence. This possibility is particularly acute when using a TA, as amplified spontaneous emission (ASE) typically leads to the central wavelength sitting on a relatively large background pedestal. 

We therefore experimentally investigated the level and effect of ASE at 689~nm in our MOPA laser system, this being the wavelength of the nearest optical transition ($^{1}$S$_{0}$-${^{3}}$P$_{1}$) to the clock transition. Fig.~\ref{fig:taCharacterisation}~(b) shows a typical spectrum of the clock laser after the 10-m PM fiber where a narrow line at 698~nm is superimposed onto a broadband pedestal measured by a homemade grating spectrometer with a resolution of 0.25~nm. Using this spectrum and with our typical beam intensity of $I= 25 $~W/cm$^{2}$, the scattering rate induced by the residual ASE at 689~nm can be estimated at $\Gamma_{r}$ = 3.6~s$^{-1}$. Given that our interferometer pulses are typically of $\sim$~ms duration, this would produce a negligible amount of scattering. 



In order to experimentally confirm the lack of off-resonant single-photon scattering, several tests have been performed at both short ($\approx$ms) and long ($\approx$s) time scales (as shown in Fig.~\ref{fig:bandpass}). These tests were performed with and without the presence of a bandpass filter ($\lambda$=700~nm, FWHM=10~nm) in the interferometry beam path, which would further reduce the intensity at 689~nm by a factor of 100. In the first measurement, we observed the Rabi oscillations on the clock transition with and without the bandpass filter shown in Fig.~\ref{fig:bandpass}~(a). It can be clearly seen that the Rabi oscillations for the two cases are mutually consistent, from which we infer that, during a ms-long interferometry pulse, spontaneous emission is negligible. The result is also confirmed by a second measurement, in which contrast of the atom interferometer as a function of interferometer time have been repeated with and without the bandpass filter and for which no significant difference is observed.

To further confirm this effect at a longer time scale, we measured the lifetime of atoms held in a static lattice whilst being simultaneously illuminated with the interferometry laser. In this case, however, it should be noted that the frequency of the laser is tuned away from the clock transition, so as to prevent any excitations to the ${^{3}}$P$_{0}$ state. This is necessary because although the collisional cross-section between ground state atoms is extremely small for $^{88}$Sr, this is no longer the case when these bosonic atoms are in the excited state~\cite{lisdat2009collisional}, which would deleteriously effect the lattice lifetime. Again the measurements were made with and without the presence of the bandpass filter and again the bandpass filter was shown to produce no discernible difference. As the data in Fig.~\ref{fig:bandpass}~(b) shows, the lattice decay time constants are independent of whether the filter is used or not. Instead, the lattice lifetime is here mainly limited by the ambient vacuum pressure~\cite{zhang2016trapped, Poli2011}. 

As the estimated scattering rate and the above tests indicate, the effect of the residual ASE at 689~nm of the MOPA output on the performance of our interferometers can be neglected at our instrumental precision. Moreover, the contribution is estimated to remain negligible even if intensities several orders of magnitude are delivered. This is because the Rabi frequency is also increased at higher intensities and will remain on a much faster timescale than any expected spontaneous scattering rate.



\subsubsection{Fiber-noise cancellation}

 \begin{figure}
    \centering
    \includegraphics[width=\linewidth]{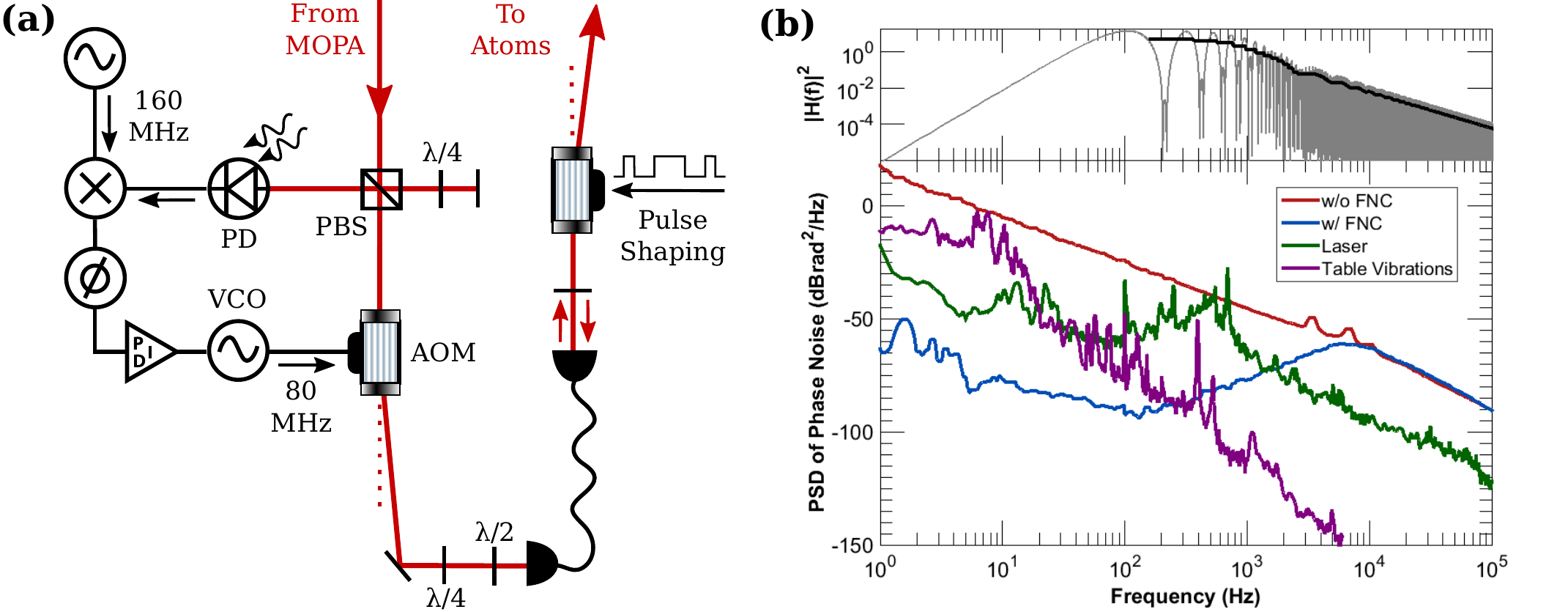}
    \caption{a) The optical path noise introduced by the optical fibre is reduced by using a fiber-noise cancellation arrangement. A small fraction of the light is picked off a reference whilst the rest is first-order diffracted by an AOM into the fibre. A window reflects a small proportion back through the fibre and the AOM, producing a beat note at 160~MHz which is used to generate an error signal and control the VCO providing the RF to the AOM. b) This setup dramatically reduces the in-loop PSD of the phase noise introduced by the optical fibre (blue line, without FNC; red line, with FNC). The dominant noise contributions at low frequencies are now from the phase noise of the laser (green line) and the vibrations of the table (purple line). The upper panel shows the calculated transfer function for this experiment (grey line), with the mean over a single oscillation period shown to make the trend clear (black line).}
    \label{fig:noiseSources}
\end{figure}

As discussed earlier, due to the sensitivity of a single-photon gravimeter to the laser phase, it is very important that the laser phase noise is as low as possible. This includes not just the phase noise of the laser itself, but also any phase noise accumulated along the optical path, such as from vibrations. For optimum performance, these sources need to be either isolated against or suppressed.

One large source of phase noise which can be efficiently suppressed is the noise introduced by the 10-m PM fiber which delivers the light from the MOPA system to the vacuum chamber. This fiber, as mentioned above, is necessary to improve the mode quality of the light and thus provide a more constant phase reference across the atomic cloud. To remove the additional phase noise introduced by the fibre, we implement a fiber-noise cancellation (FNC) apparatus. Here a standard optical interferometer configuration was adopted~\cite{ma1994delivering, riehle2017optical}, as depicted in Fig.~\ref{fig:noiseSources}~(a). The interferometry light is first split into a local reference beam and an outgoing beam on a polarising beamsplitter. The outgoing beam passes through an 80~MHz AOM and the first-order diffracted light is coupled to the optical fiber. At the end of the fiber, some of the light is reflected back through the fiber by a window located just after the fiber output. This light passes back through the AOM and the first diffracted order is compared to the reference beam by monitoring their beatnote as measured by a photodiode. The error signal acquired by beating the 160~MHz beatnote with a reference RF signal at 160~MHz from a DDS is then used to modulate the voltage-controlled oscillator (VCO) that drives the AOM, where we employ a dedicated servo amplifier with a bandwidth of order 10~kHz. Note that since the light passes through both the fiber and the AOM twice, the light is stabilized at both ends of the fiber. Fig.~\ref{fig:noiseSources}~(b) shows that we obtain a reduction of phase noise of about 50~dB up to the Fourier frequency of 400~Hz and suppress this noise beneath the noise of the laser itself, up to frequencies of around 2~kHz, which therefore becomes the technological limit for our single-photon gravimeter. This is highlighted by looking at the calculated transfer function for our experimental parameters ($\tau$=450~$\mu$s, $T=4$~ms), which shows that the laser noise dominates in the most critical frequency range (Fig.~\ref{fig:noiseSources}~(b)).

In this configuration, to operate the gravimeter with the FNC setup, we added a second AOM for shaping the interferometer pulses~\cite{falke2012delivering}, but otherwise kept the number of optical components to a minimum. This is because any further components will introduce noise uncompensated for by the FNC setup. To characterise this noise we measured the vibrations of our optical table and the vibrations of a mirror mount attached to the table, using a seismometer for low frequencies and a piezo accelerometer for high. The vibrations on the mirror mount were seen to follow those of the table, as expected for them behaving as a rigid body. The PSD of these vibrations substantially exceeds the noise FNC setup at low frequencies, showing that any components added after the cancellation will degrade the performance of the interferometer and must therefore be avoided or vibrationally isolated, such as is the case for the retro-reflecting mirror in multi-photon accelerometers~\cite{zhou2012performance, hensley1999active,tang2014programmable}.

\subsection{Lattice launch}\label{subsec:latticeLaunch}

 \begin{figure}
        \centering
        \includegraphics[width=1\linewidth]{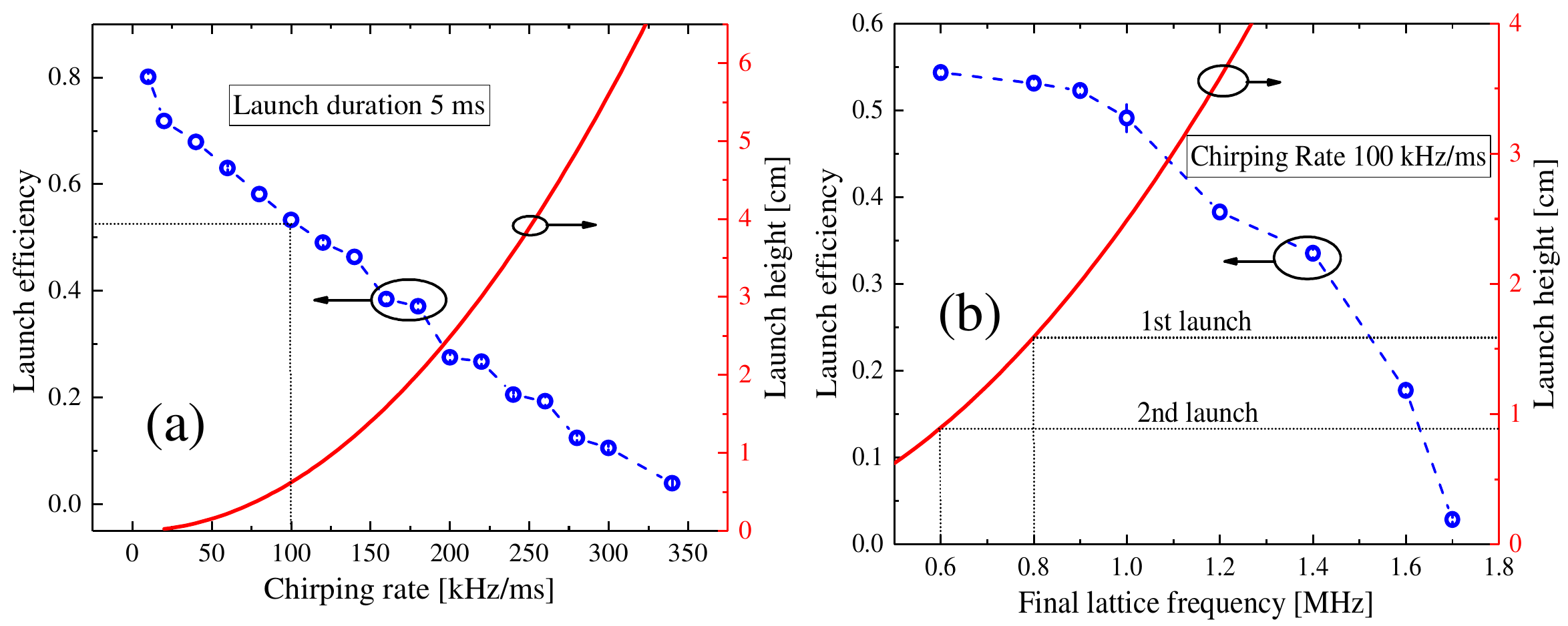}
        \caption[Green lattice launch performance.]{Lattice launch efficiencies and calculated launch heights as a function of the frequency chirping rate (a) and the final lattice frequency detuning (b). In order to approximately evenly distribute the atoms into the two arms of the interferometer, launch efficiencies close to 50\% are selected, meaning we typically operate at a chirping rate close to 100~kHz/ms, which corresponds to an acceleration of 2.7~$g$. We set the final lattice detuning in the range 0.5~MHz $<\delta_{L}<$ 1~MHz where the launch efficiency has a relative flat range. The final frequency differences are 0.8~MHz (8~ms) for the first launch and 0.6~MHz (6~ms) for the second launch. The resultant apogees of the two clouds are 16~mm and 9~mm above the MOT position, as indicated by the dashed lines.}
        \label{fig:launch}
\end{figure}



In order to increase the total time the atoms are in free fall and hence increase the possible interferometry time and sensitivity, the atoms are launched upwards by a moving optical lattice. The lattice is formed by a 532-nm laser with an intensity of 10$^{4}$~W/cm$^{2}$, for which we estimate a negligible scattering rate of $0.08$~s$^{-1}$, given typical transport times of $\sim$10~ms. Transport losses are therefore dominated by Landau-Zener tunnelling~\cite{peik1997bloch}, allowing an efficient launch to be achieved by using a slower chirp rate over a longer time. When operating in a gravimeter configuration, launch efficiencies of 70\% were achieved with a chirp rate of 10~kHz/ms for a duration of 65~ms.





In the case of the gradiometer configuration, however, the situation is complicated by the need to sequentially launch two clouds of approximately equal populations to two distinct heights. We therefore characterised the launch performance as a function of the chirping rate, which sets the launch acceleration, and the final frequency detuning, which determines the final velocity (Figure~\ref{fig:launch}). Lattice launch efficiencies and launch heights as a function of the frequency chirp rate for a fixed launch duration of 5~ms are shown in Fig.~\ref{fig:launch}~(a). In order to evenly distribute the atoms between the two launched clouds, a chirping rate of 100~kHz/ms was selected, which corresponds to an acceleration of 2.7~$g$ (dashed line in Fig.~\ref{fig:launch}~(a)).

Fig.~\ref{fig:launch}~(b) depicts the lattice launch efficiency as a function of the final lattice frequency difference at the selected chirp rate of 100~kHz/ms. To maintain a good launch efficiency, the final lattice detuning should be in the approximately flat region between 0.5~MHz and 1~MHz. In order to sequentially launch two clouds from a single original cloud, we set the final lattice detuning in the region 0.5~MHz $<\delta_{L}<$ 1~MHz where the launch efficiency almost remains constant. Under these conditions, we operate at a final frequency difference of 0.8~MHz for the first launch and 0.6~MHz for the second launch, resulting in respective apogee positions of the two clouds of 16~mm and 9~mm above the MOT position. Under these conditions each cloud contains $\approx3.5\times10^{5}$ atoms.

\subsection{Single-photon velocity selection} \label{subsec:velSel}

As anticipated in Section~2, to perform interferometry with a sufficiently high contrast in the presence of a low Rabi frequency for the optical clock transition, a sample with a narrow velocity distribution is required, which can be achieved by the application of a preliminary $\pi$ pulse~\cite{kasevich1991atomic, moler1992theoretical}. In order to obtain this condition, before the interferometry sequence, we performed a velocity selection from the launched clouds using the interferometry laser. This is one of the first implementations of velocity selection using an ultra-narrow optical transition and it is an essential tool for our interferometer.

This process is similar to the selection operated with Raman or Bragg transitions in the sense that the selected atoms can have an arbitrarily small velocity spread by using a long selection pulse. This statement holds true as long as the selection pulse does not exceed the excited state lifetime and the coherence time of the laser. These limits are not severe in practice for the strontium clock transition and for a 1~Hz linewidth laser. It remains true, however, that from run-to-run of the experiment there will be frequency fluctuations of the interferometry laser of the order of 100~Hz that can change the center velocity of the selected cloud. Because of this velocity fluctuation it is essential to use the same laser for velocity selection and the interferometry sequence.

To test our capability to select the atomic velocity we first apply a velocity-selective pulse on the atoms which have been released from the lattice with a vertical temperature of 700~nK. We then measure the temperature of the selected cloud by sending a second pulse with variable frequency and counting the atoms transferred by this second pulse. The velocity distribution can then be extracted from the spectroscopic data by fitting to the following expression,
\begin{equation}
\mathcal{C}(\delta)=\int_{-\infty}^{+\infty}dvG_{v}(v)P_{e}(\delta-kv),
\label{eq6}
\end{equation}
where $P_e(\delta)$,
\begin{equation}
P_{e}(\delta)=\frac{{\Omega}^{2}}{{\Omega}^{2}+\delta^{2}}\sin^{2}\left(\frac{\Delta t}{2}\sqrt{{\Omega}^{2}+\delta^{2}}\right),
\label{eq4}
\end{equation}
is the probability for an atom to be found in the excited state when laser light with detuning $\delta$ and Rabi frequency $\Omega$ is shone for a time $\Delta t$. The function $G_v(v)$ describes the atomic velocity distribution and we assume this to be represented by a Gaussian:
\begin{equation}
G_{v}(v)=\frac{A}{\sqrt{2\pi}\sigma_{v}}\exp\left(\frac{-(v-v_{0})^{2}}{2\sigma_{v}^{2}}\right),
\label{eqgauss}
\end{equation}
where $\sigma_v$ is the Gaussian width, $v_0$ the mean velocity and $A$ is a constant that is proportional to the total atom number.



\begin{figure}
    \centering
    \includegraphics[width=1\linewidth]{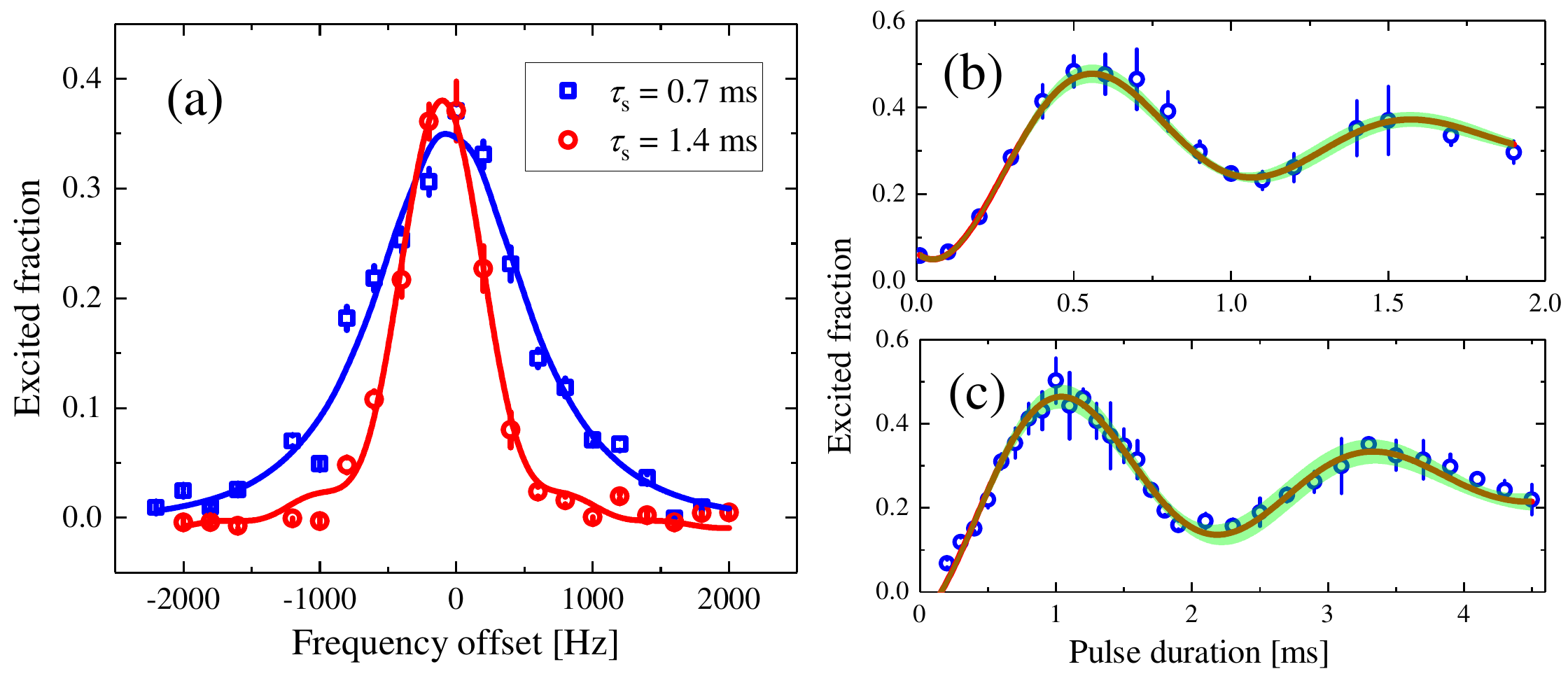}
    \caption{(a) Excited fraction of atoms on the clock transition for two different velocity selection pulse durations: 0.7~ms (blue squares) and 1.4~ms (red circles) at $1/e^{2}$ beam radius of the clock beam $r_{0}=0.45$~mm. The solid curves represent the best fit with the function $C(\delta)$ (see text), from which momentum widths of 0.044~$\hbar k$ and $0.023\hbar k$ are estimated, respectively. (b) and (c) Excited fraction as a function of interaction time (Rabi oscillations) for $r_{0}=0.45$~mm (b) and $r_{0}=0.90$~mm (c), respectively. The red line represents the best fit of the data with a damped sinusoidal function. The green shadow represents a 95\% confidence band of the fit.} 
    \label{fig5}
\end{figure}


In order to guarantee a sufficient signal-to-noise ratio at detection, we set the pulse duration of the spectroscopic pulse to be $\Delta t=1.4$~ms, resulting in a minimum measurable momentum of 0.024~$\hbar k$. Figure~\ref{fig5}~(a) shows clock transition spectra recorded following initial square, velocity-selection pulses with durations of 0.7~ms and 1.4~ms. The fit of the two datasets to Eq.~\ref{eq6} indicates a momentum width of 0.044~$\hbar k$ and 0.023~$\hbar k$ for selection pulse durations of 0.7~ms and 1.4~ms, respectively, with corresponding temperatures of 860~pK and 235~pK. However, by extending the selection pulse duration from 0.7~ms to 1.4~ms, the detected atom number is decreased from $10^{4}$ to $5\times10^{3}$.




By instead applying a second pulse of fixed frequency and tunable duration, the output will produce Rabi oscillations between the $^{1}\text{S}_{0}$ and $^{3}\text{P}_{0}$ states. Figs.~\ref{fig5}~(b) and (c) show the measured Rabi oscillations as a function of laser pulse duration and are used to determine the durations of the $\pi$ and $\pi/2$ pulses for the interferometry sequence. The oscillations are recorded with a typical value of the static magnetic field $B = 350$~G and at laser peak intensities of 25~W/cm$^{2}$ (Fig.~\ref{fig5}~(b)) and 6~W/cm$^{2}$ (Fig.~\ref{fig5}~(c)). The experimental results fit well with a damped sinusoid with a corresponding Rabi frequency of $2\pi\times1013(22)$~Hz ($2\pi\times436(12)$~Hz) and a damping time of $0.87(5)$~ms ($2.35(33)$ ms) shown in Fig.~\ref{fig5}~(b) ((c)). This small damping time can be explained by the expansion of the cloud in the horizontal direction and the small beam size of the interferometry laser beam, which is comparable to the atomic cloud size in the horizontal direction.  

 

\subsection{Interferometer contrast \& visibility}




The fringe visibility of an atom interferometer is highly dependent upon the spatial extent of the atomic source in comparison to the size of the interferometry beams. This is because the output of the interferometer is an average measurement across the whole atomic population, meaning that any phase variations across the cloud will tend to lead to a washing out of the interference fringes. These variations can arise from several sources, including either directly from variations in the wavefront of the interferometry beam or from variations in the intensity profile leading to Rabi frequencies that are dependent upon atom position. We investigate these effects on our interferometer by looking in both the longitudinal (vertical) and transverse (horizontal) directions relative to the beam.


 \begin{figure}
    \centering
    \includegraphics[width=1\linewidth]{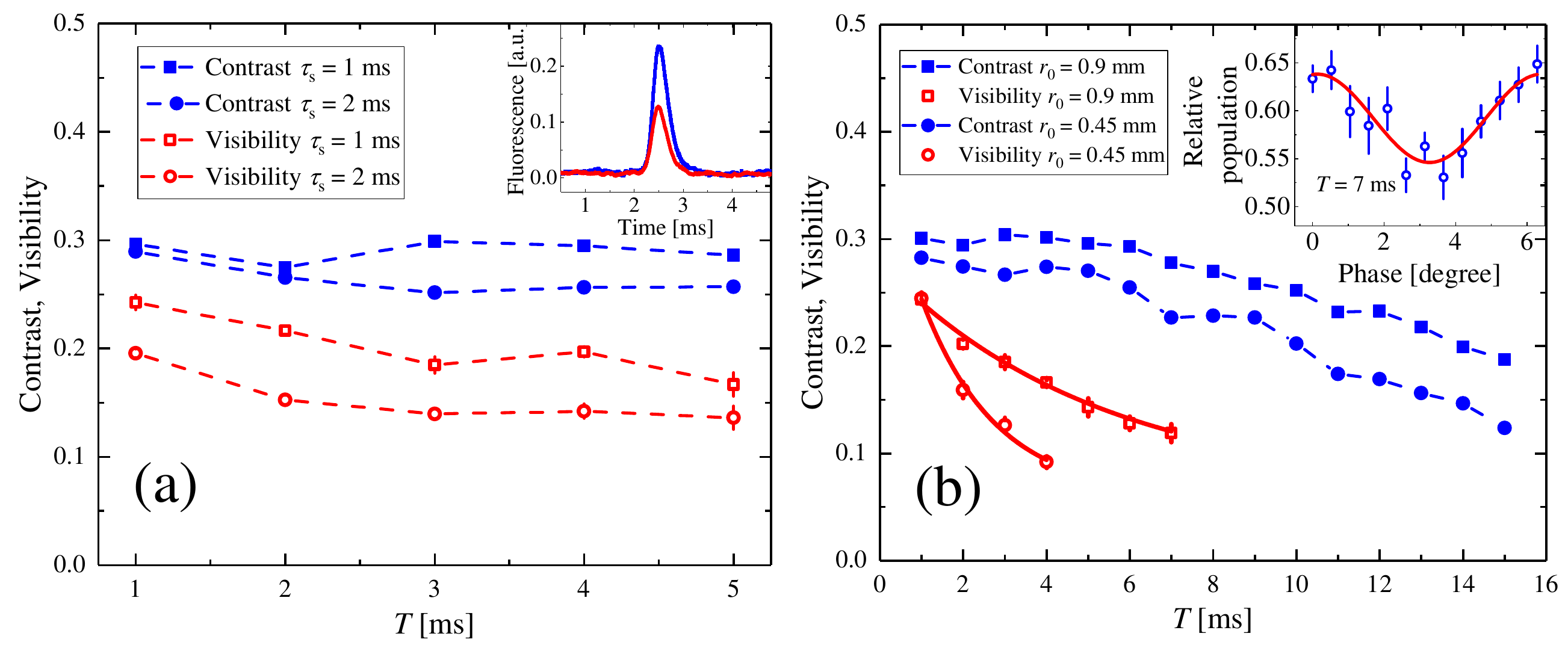}
    \caption{(a) Measured interferometer contrast and fringe visibility as a function of the interferometry time $T$ following a selection pulse duration of 1~ms (squares) and 2~ms (circles). The inset shows fluorescence signals of the selected cloud passing through the detection system (blue curve for 1~ms and red curve for 2~ms). (b) Measured interferometer contrast and fringe visibility as a function of time $T$ for different $1/e^{2}$ interferometry beam radii: $r_{0}=0.45$~mm (circles) and $r_{0}=0.9$~mm (squares). From the fit of the visibility with an exponential decay curve, we obtain decay times of 1.7(6)~ms and 5.2(9)~ms, for the small and large beams, respectively. The inset depicts an example of the observed fringe at time $T= 7$~ms with a beam radius of 0.9~mm (each point is the average of 10 shots). }
    \label{fig:contrastVisibility}
\end{figure}

The expansion rate of the cloud, and therefore its spatial extent, can be controlled in the vertical direction by employing the velocity selection techniques discussed above. The dependence of the contrast and fringe visibility on the vertical velocity distribution is shown in Fig.~\ref{fig:contrastVisibility}~(a).  While we observed a lower \emph{fringe visibility} for the shorter selection pulse, the contrast is almost unchanged. This is because the fringe visibility strongly depends upon the phase noise of the interferometry laser and the detection efficiency, whilst the contrast mainly relies upon the fraction of atoms which can be stimulated to undergo the optical transition. The main reason for this fringe visibility reduction is the decreased number of atoms when using the longer selection pulse duration, resulting in a lower signal-to-noise ratio of the detection signal. Here we observed that the detected fluorescence is reduced by 50\% by increasing the selection pulse duration from 1~ms to 2~ms (inset of Fig.~\ref{fig:contrastVisibility}~(a)). As utilising a narrower vertical velocity class diminishes the visibility, this demonstrates that in this regime the dominant form of this particular form of visibility loss arises from the atom number and not from the efficiency of the interferometry pulses themselves.




The horizontal temperature cannot be controlled as easily as in the vertical case, so in order to investigate the effect of the horizontal temperature, we instead altered the interferometry beam size, whilst keeping a constant intensity of 6~W/cm$^2$, and observed the contrast and fringe visibility as a function of interferometry time (Fig.~\ref{fig:contrastVisibility}~(b)). In performing these tests, we achieve a total interferometry time of 30~ms, which is longer than has previously been reported~\cite{hu2017atom}.

We perform experiments with a beam radius ($1/e^{2}$) of either $r_{0}=0.9$~mm or reduced to $r_{0}=0.45$~mm (Fig.~\ref{fig:contrastVisibility}). The effect of the beam size on visibility is characterised by fitting the measured visibility to an exponential decay. We obtain a decay time of $5.2 (9)$~ms for $r_{0}=0.9$~mm, which is about three-times better than the decay time of $1.7(6)$~ms observed in the small beam size case of $r_{0}=0.45$~mm. This corresponds to an equivalent threefold increase in the sensitivity of interferometers. The reduction in contrast, however, is much less pronounced and the difference in the decay rate is not as obvious, though the contrast does again appear to decay faster for the smaller beam. We attribute these observations to the transverse thermal motion of the atoms, which have a temperature in these dimensions of approximately 700~nK. The resultant thermal expansion means that atoms will interact with different regions of the interferometry beam during each interferometry pulse, introducing dephasing due to the effective presence of many different Rabi frequencies.

These effects are obviously lessened the larger the beam is compared to the cloud, as observed above, and for lower temperatures. Expanding the size of the interferometry beam is in principle trivial, though care must be taken to maintain an intensity high enough to produce sufficiently fast Rabi oscillations and also to ensure the beam is considerably smaller than its entrance aperture, if damaging diffraction effects are to be avoided. Both of these considerations limit what is achievable with our current apparatus. Lower temperatures can be attained by introducing additional cooling stages and moving towards quantum degeneracy~\cite{stellmer2009bose,stellmer2013production}, although this generally increases the duty cycle of the experiment and so decreases the device sensitivity.






\subsection{Gravimeter Sensitivity}


The interferometer was operated as a gravimeter, running a Mach-Zehnder sequence on a single, free-falling atomic cloud (see Section~\ref{subsec:gravProceduce}). The sensitivity of the interferometer is calculated from the inferred value of $\delta g/g$, as determined by measuring the phase fluctuations $\delta\Phi$ at the slope of the central fringe~\cite{hu2017atom, mazzoni2015large}:
\begin{equation}
\frac{\delta g}{g}=\frac{\delta\Phi}{\frac{\omega_a}{c}gT^{2}\left[1+\left(2+\frac{4}{\pi}\right)\frac{\tau_R}{T}+\frac{8}{\pi}\left(\frac{\tau_R}{T}\right)^{2}\right]}.
\label{eq:gravimeterNormPhase}
\end{equation}



\begin{figure}
    \centering
    \includegraphics[width=\linewidth]{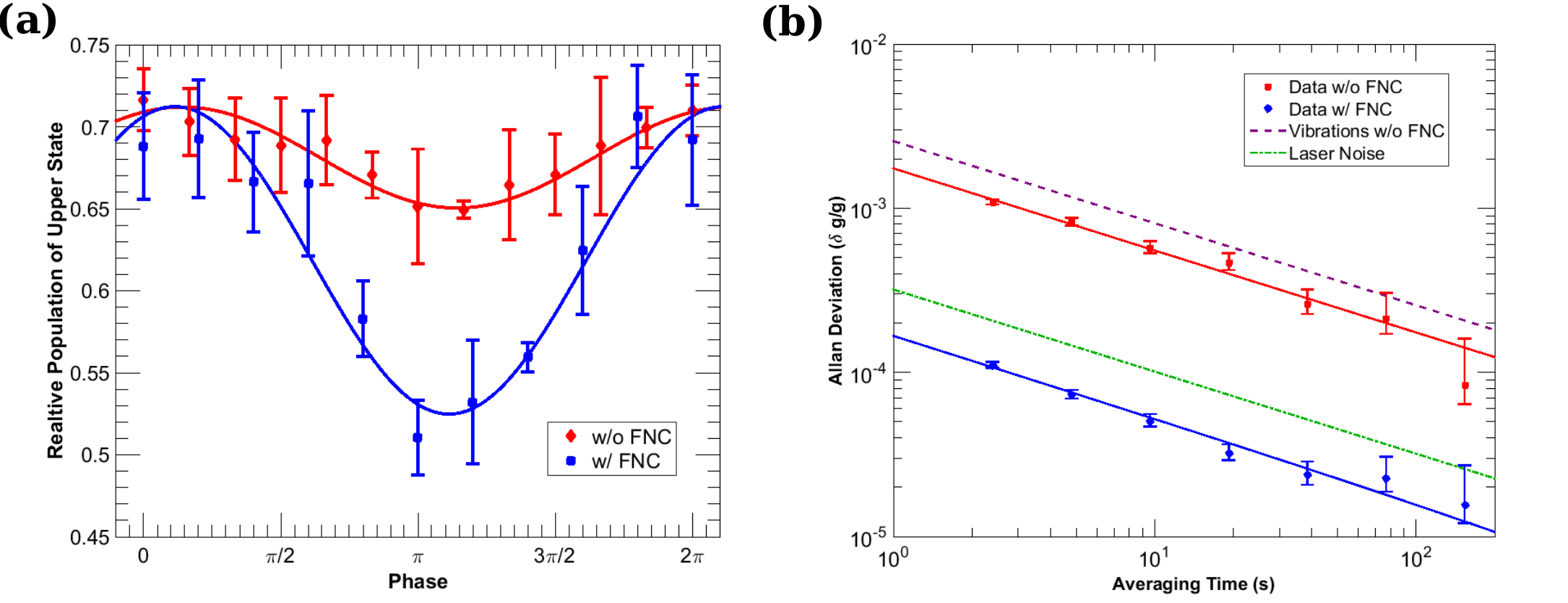}
    \caption{a) The observed change in the fringe visibility with the FNC setup (blue squares, $\approx 19$~\% visibility)  and without FNC setup (red diamonds, $\approx7$~\% visibility). Each point is the average of 10 measurements and the curves are fits to a sinusoid. b) Allan deviation of gravitational acceleration measurements performed with the single-photon gravimeter for an interferometer time $T= 4$~ms. The best fractional acceleration stability, obtained with the FNC setup, is $2.1\times10^{-4}$ at 1~s integration time, averaging down to $1.7\times10^{-5}$ after 150~s (blue curve). This value is a factor of 5 better than the stability obtained with a gravimeter without the implementation of the FNC setup (red curve). Also shown in the figure are the estimated effects due to the phase noise induced by the 10-m PM fiber without the FNC setup (dashed purple line) and the noise of the clock laser beam (dash-dotted green line). The limits expected from the shot noise and the noise induced by the fibre with the FNC operational are far below the measured data and are therefore not shown.}
    \label{fig:gravMeas}
\end{figure}

The short-term sensitivity is characterized by the Allan deviation, which is shown for a single-photon gravimeter with an interferometry time $T=4$~ms in Fig.~\ref{fig:gravMeas}. Measurements were performed using the smaller beam radius ($r_{0}=0.45$~mm) both with and without the FNC operating; as a comparison and also in order to allow the various noise sources to be investigated. We observed for both cases that the Allan deviations at short times scale as the inverse square root of the integration time, as anticipated.

Without the FNC, the measured sensitivity is $1.1 \times 10^{-3}$ at 1~s of averaging time and $9\times 10^{-5}$ at 150~s of averaging time. However, since the FNC is not activated, the atomic interference fringes are quickly washed out. This is particularly clear by comparing the fringe visibility: $\approx19$~\% with the FNC setup and $\approx 7$~\% without (Fig.~\ref{fig:gravMeas}~(a)).  Introducing the FNC setup leads to an improvement in performance by an approximate factor of 5: $2.1\times 10^{-4}$ at 1~s of integration time and averaging down to $1.7\times 10^{-5}$ after 150~s. This highlights that the stabilization of the phase noise introduced by the 10-m PM fiber is of crucial practical importance in reducing the phase noise imparted onto the atoms.

These data can be compared to the expected sensitivity limits imposed by the various noise sources (Fig.~\ref{fig:noiseSources}) by use of the sensitivity function method~\cite{cheinet2008measurement,le2008limits}, by inputting the measured PSDs of these noise sources into Eq.~\ref{eq:transferFunction}. As implied from the improved gravimeter performance following the implementation of the FNC setup, the gravimeter appears to be limited by the noise introduced by the 10-m PM fiber when operated without the FNC, with the calculation showing a reasonable agreement with the data (Fig.~\ref{fig:gravMeas}~(b)). In the case where the FNC is operational, the gravimeter instead appears to be limited by the phase noise of the laser itself, with the limit calculated from the measured PSD being in reasonable agreement (factor of 2) with the observed data. For comparison, the noise limits expected from both the noise of the FNC system and the shot noise ($10^{4}$ atoms, visibility of 19~\%) are around an order of magnitude lower, respectively being $1.4\times10^{-5}/\sqrt{\tau}$ and $1.7\times10^{-5}/\sqrt{\tau}$.

We therefore conclude that primary limitation to the performance of the single-photon gravimeter presented here is the interferometry laser itself. However, we note that this is not a fundamental limitation and, as calculated earlier (Fig.~\ref{fig:gravSensitivityCalc}), it can be surpassed by employing state-of-the-art optical clock lasers~\cite{matei201715, zhang2017ultrastable}. In such a situation, we anticipate that the shot noise limit should be approachable.

\subsection{Gradiometer characterization}
\subsubsection{Artificial phase shift}

With a vertical separation of $\Delta r_{0}\approx$ 1.9~mm and velocity difference $\Delta v_{0}\approx 6.5$~mm/s between two accelerometers at the beginning of the gradiometer, the two leading phase shift contributions, coming from the height separation and initial velocity difference, are $\approx10^{-7}$~rad and $\approx10^{-6}$~rad, respectively. They are much smaller than the atom-shot-noise-limited phase resolution of $\approx1/\sqrt{10^4}=10$~mrad for $10^4$~atoms. Therefore, when plotting the signal of the upper accelerometer against the lower one, the dataset will produce a closed Lissajoux figure. Several schemes have been proposed to tune the ellipse angle by introducing additional external fields, but all these field values are usually either difficult to calibrate or to realize~\cite{delaguila2018bragg, foster2002method}.  Here we achieve a differential phase shift between two atom interferometers by simply adjusting the phase difference $\delta\phi$ between the two RF signals used to control the delivered beam frequencies~\cite{hu2017atom}.

\begin{figure}
    \centering
    \includegraphics[width=1\linewidth]{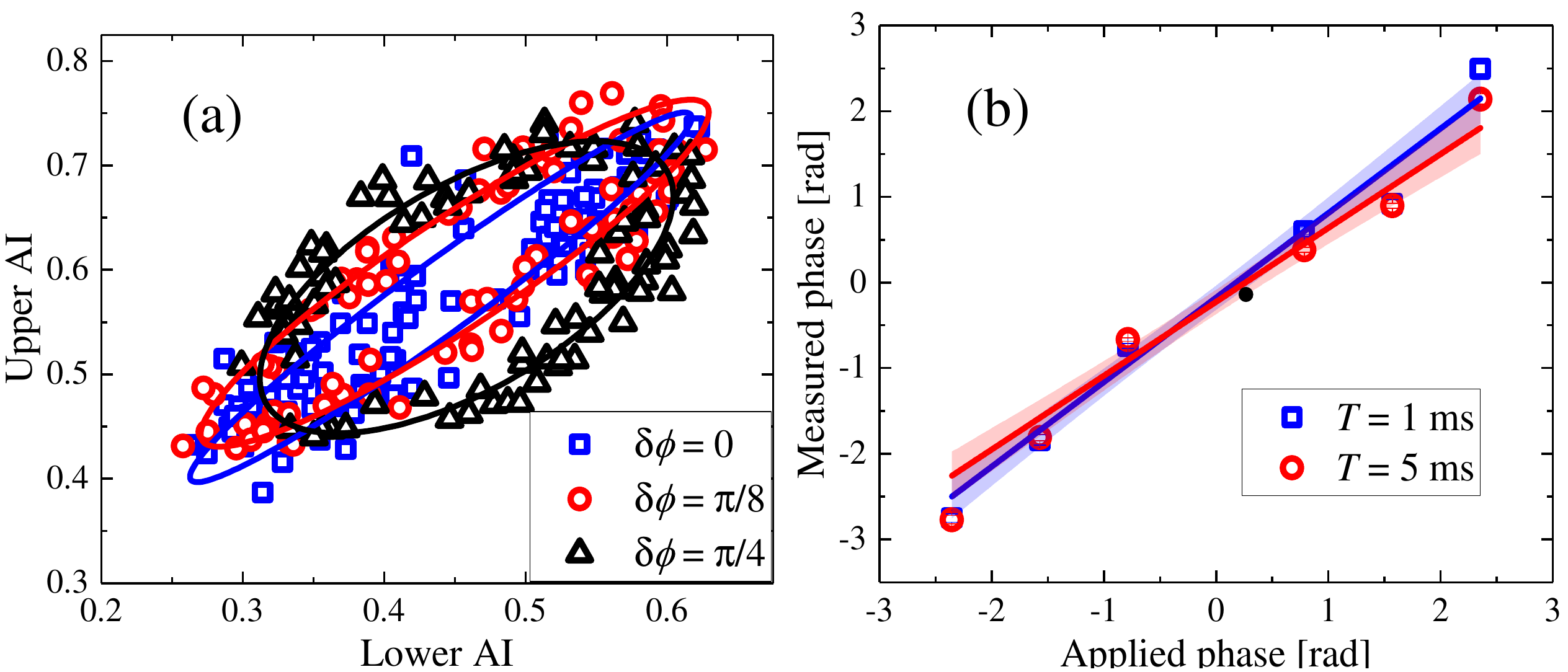}
    \caption{(a) Measured relative populations for the upper and lower atom interferometers for different relative phase shifts at $T=5$~ms. The least-squares ellipse fitting (solid curves) gives a relative phase $0.21(5)$~rad (blue curve), $0.38(4)$~rad (red curve) and $0.90(3)$~rad (black curve) for the applied relative phase shifts of $\delta \phi =0$~rad (blue squares), $\delta \phi =\pi/8 $ rad (red circles) and $\delta \phi =\pi/4$~rad (black triangles), respectively. By increasing the applied relative phase shifts, the ellipse is made to open progressively, illustrating the appearance of the expected fixed relative phase between two atom interferometers. (b) The measured phase shifts as a function of different applied phase shifts $\delta\phi$ at two different interferometer times: $T$ = 1~ms (blue squares) and $T$ = 5~ms (red circles). Each dataset is fitted by a linear function (solid lines), indicating a slope of 0.98(8) for $T=1$~ms and 0.86(9) for $T=5$~ms. The shaded area indicates the 1-$\sigma$ confidence interval of the fit. }
    \label{fig10}
\end{figure}


To demonstrate the artificial phase shift $\delta\phi$ appearing in the gradiometer, we tune $\delta\phi$ and measure the ellipse angle by least-squares ellipse fitting. As shown in the inset of Fig.~\ref{fig10}~(a), when the artificial phase shift $\delta\phi$ is increased, the ellipse is seen to progressively open, illustrating the expected fixed relative phase between the two atom interferometers.  Note that when the applied artificial phase shift $\delta\phi$ is non-existent ($\delta\phi$=0~rad), the measured phase shift is 0.21(5)~rad. This inconsistency is due to a systematic error which occurs from the ellipse fitting technique when the opening angle approaches 0~rad~\cite{foster2002method}. This error can be largely reduced by tuning the ellipse angle to near $\pm\pi/2$~rad, a potential advantage of introducing such an artificial phase shift.



To further verify the robustness of this technique at different phase noise conditions, we varied the artificial phase shift $\delta\phi$ and measured a series of ellipse angles for two different interferometer times: $T=1$~ms and $T=5$~ms (Fig.~\ref{fig10}~(b)). Here, we find a good agreement of the measured phase shifts and the applied artificial phase shifts with a slope of 0.98(8) for $T=1$~ms and 0.86(9) for $T=5$~ms. The slight offset from the ideal condition with a slope of 1 and the difference between the two cases are mainly due to the ellipse fitting performance and detection noise from the photon collection efficiency at the output of the gradiometer~\cite{hu2017atom,delaguila2018bragg}. 


\subsubsection{Gradiometer sensitivity}
\begin{figure}
    \centering
    \includegraphics[width=0.55\linewidth]{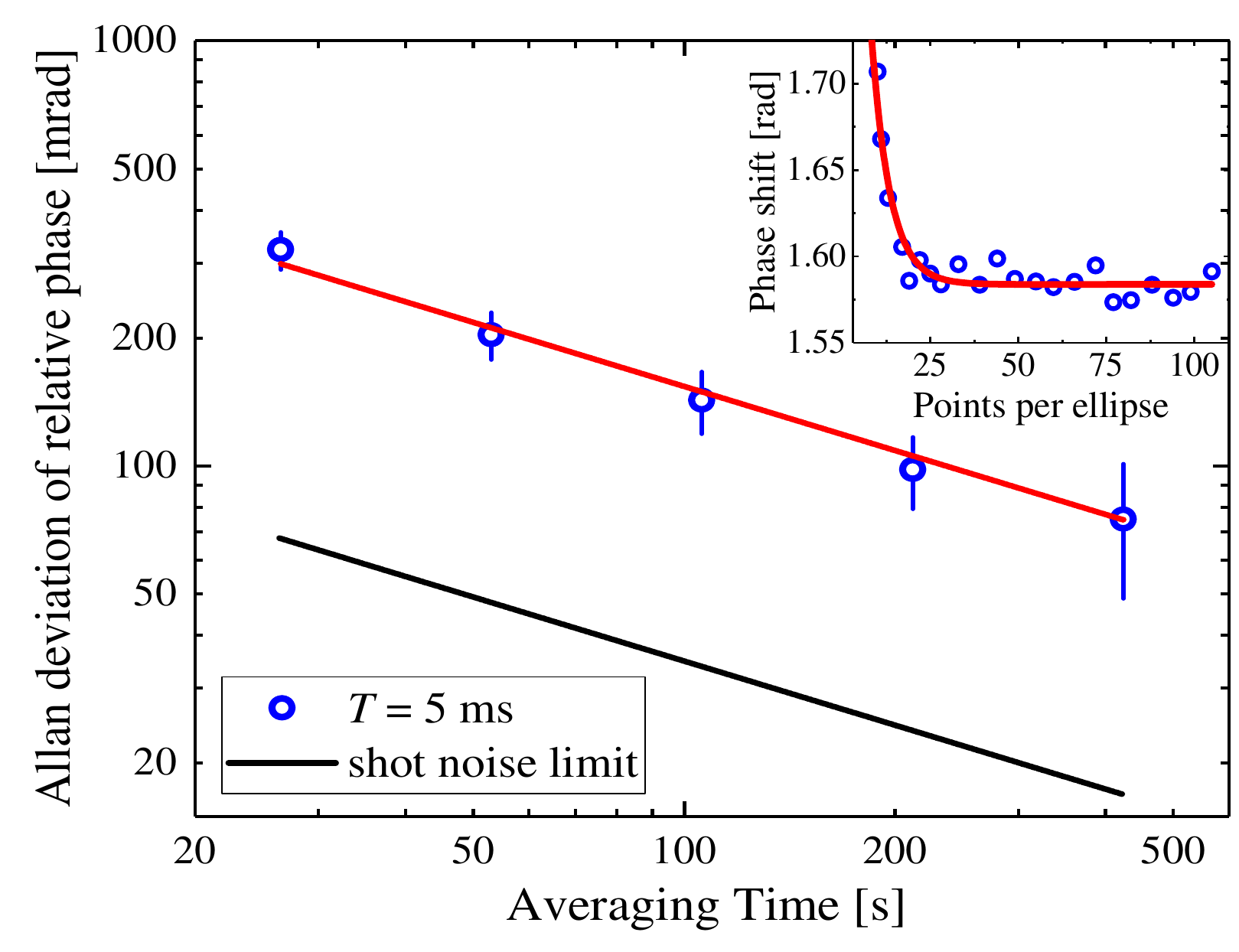}
    \caption{Allan deviation of the relative phase shift for the gradiometer for $T=5$~ms at an artificial phase shift of $\delta\phi=\pi/2$~rad. The Allan deviation scales down as the inverse of square root of the averaging time $\tau$, as shown by the fit to the data (red solid curve), indicating a sensitivity of 76.5~mrad at 400~s of integration time. The inset shows the average ellipse angle determined by the fitting method as a function of the number of points used in the ellipse. The red solid curve represents the results fitted by an exponential decay wave, indicating more than approximately 5 data points are needed to accurately obtain the applied phase shift $\delta\phi=\pi/2$~rad.}
    \label{fig18}
\end{figure}


To test the sensitivity of our apparatus using the method of adding an artificial phase shift studied above, we observed the statistical fluctuations of the gradiometer measurements at the most sensitive point of the phase noise $\delta\phi=\pi/2$~rad and with $T= 5$~ms. The measurements have been repeated with a set of $1000$~cycles. The cycle time was set to 2.4~s, resulting in an overall acquisition time of approximately 40~minutes. For the ellipse fitting, the systematic error of the differential phase measurement also depends on the number of points per ellipse. To investigate this effect, we divided the data into a series of consecutive data points.  The inset of Fig.~\ref{fig18} shows an analysis of the measured phase shifts as a function of the number of points per ellipse. The results are fitted by an exponential decay function and indicate that more than approximately 5 data points are needed to achieve a low systematic error. Fig.~\ref{fig18} shows the Allan deviation plot for 11 points per ellipse. As expected for the cancellation of the common-mode laser phase noise, the Allan deviation scales with the inverse square root of the averaging time ($1.53/\sqrt{\tau}$~rad), illustrating that the main noise contribution comes only from white phase noise, and results in a value of 76.5~mrad at $\tau =$~400~s. The measured sensitivity is therefore similar to the one in Ref.~\cite{hu2017atom} and is about five times worse than the shot-noise-limited sensitivity estimated for $\approx10^4$ atoms and with a typical contrast of $\approx30$\%.

\subsection{Towards single-photon atom interferometers with $^{87}$Sr atoms} \label{subsec:sr87}

  \begin{figure}
    \centering
    \includegraphics[width=\linewidth]{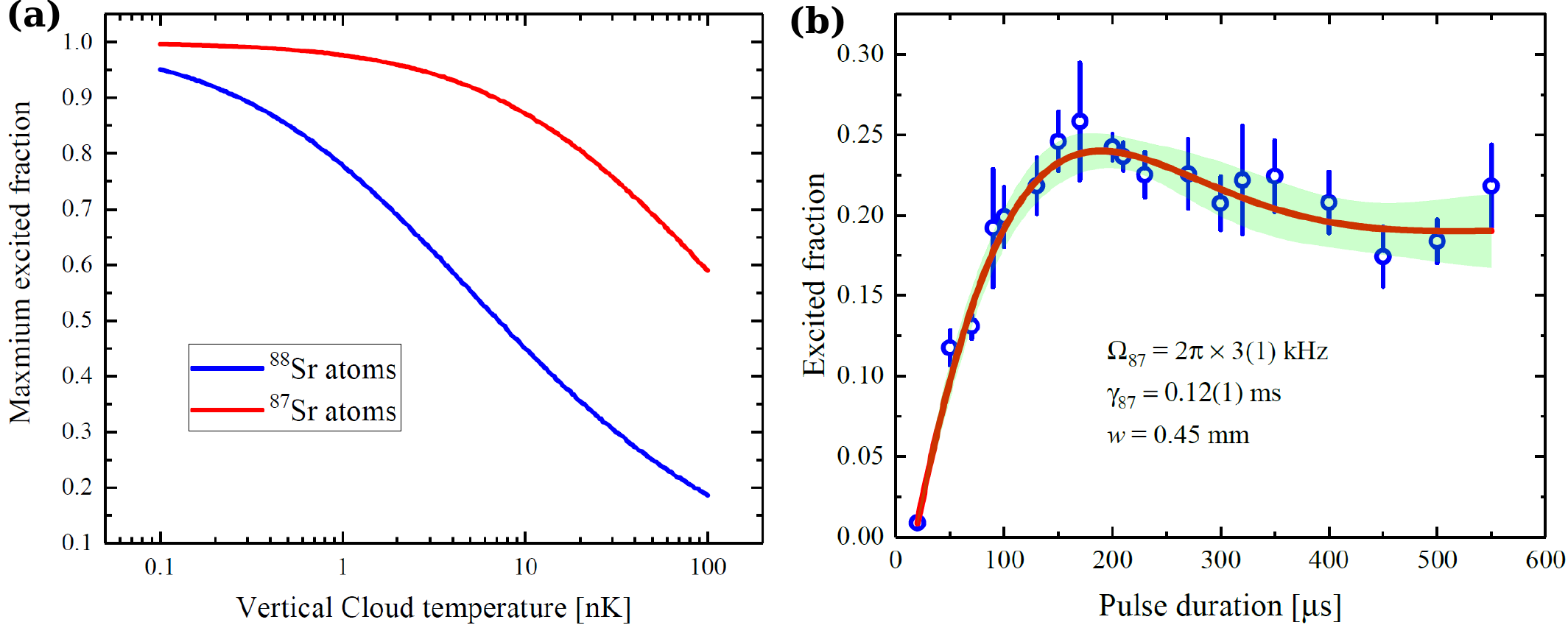}
    \caption{a) Estimated maximum excited fraction of atoms as a function of the vertical cloud temperature and for a fixed horizontal temperature of 700~nK. The calculation is performed for $^{88}$Sr atoms (blue curve) with a magnetic field $B=350$~G and for $^{87}$Sr atoms (red curve). The interferometry laser beam parameters are radius $r_{0}$ = 0.45~mm and power $P=80$~mW. b) Measured Rabi oscillation in a freely-falling cloud of $^{87}$Sr atoms with a clock laser $1/e^{2}$ beam radius of $r_{0}=0.45$ mm and power of $80$ mW. The excited fraction of atoms as a function of the interferometry laser pulse duration is fitted by a damped sinusoid function, obtaining a Rabi frequency of $\Omega_{87}=2\pi\times3(1)$~kHz and a decay time of $\gamma_{87}=0.12(1)$~ms (the green shaded region indicates the 95\% confidence band).}
     \label{fig:sr87}
\end{figure}

As discussed in Section~\ref{subsec:rabiFreq}, the fermionic isotope $^{87}$Sr has potential advantages compared to the bosonic $^{88}$Sr, with one such advantage being a higher Rabi frequency for similar intensities, assuming a realistic magnetic field is applied in the case of $^{88}$Sr. A high Rabi frequency should lead to more efficient pulse transfers and we simulate the pulse efficiency on an atomic cloud with different temperatures in order to study this effect.

The calculation is performed with the following experimental parameters: interferometry laser $1/e^2$ beam radius of $r=0.45$ mm and power $P=80$~mW; the atomic cloud having an initial FWHM in the horizontal (vertical) dimension of 300~$\mu$m (70 $\mu$m) and with a fixed horizontal temperature of 700~nK; and a magnetic field of 350~G for $^{88}$Sr. Other decoherence mechanisms such as limited excited state lifetime, cold collisions and atom losses have been neglected. Fig.~\ref{fig:sr87} gives the maximum excited fraction of atoms for different vertical cloud temperatures for the two different isotopes and clearly shows that the excited fraction of atoms has a significant dependence on the vertical cloud temperature and that a higher excited fraction can be attained with $^{87}$Sr. For instance, with a vertical temperature of 10~nK, the maximum excited fractions are 87\% and 45\% for $^{87}$Sr and $^{88}$Sr atoms, respectively. This shows that achieving a high excitation fraction with $^{88}$Sr requires additional cooling down to ultra-cold regimes.





We have performed preliminary Rabi oscillation measurements using free-falling $^{87}$Sr, achieving a Rabi frequency of $\Omega_{87} =2\pi\times3(1)$~kHz (Fig.~\ref{fig:sr87}). The atoms first undergo velocity selection resulting in a vertical temperature of $\approx$1~nK. However, the horizontal temperature is considerably higher than in the case of $^{88}$Sr, being $\approx$12~$\mu$K and therefore a factor of 20 higher. We attribute both the reduced efficiency compared to our calculations and the extremely large observed damping (damping time of 0.12(1)~ms) on this large horizontal temperature. The optimization of the $^{87}$Sr MOT cloud in our apparatus remains the scope of future work.
\section{Conclusions}

In conclusion, we have characterized the performance of both a gravimeter and a gravity gradiometer based on single-photon transitions with freely falling strontium atoms. Through the use of a laser injection-lock scheme, we have demonstrated a suitable MOPA laser source at 698~nm with a usable power of up to 80~mW. We have shown that the resonant part of the ASE on the $^{1}\text{S}_{0}$-$^{3}\text{P}_{1}$ intercombination transition at 689~nm induces negligible decoherence on the experimental fringe visibility and contrast at our experimental parameters and precision. We have experimentally investigated the possibility of optimizing the contrast and fringe visibility by adjusting various factors, including the atomic cloud velocity distribution and the interferometry beam size. 

Our gravimeter has a fractional stability of $1.7\times10^{-5}$ after 150~s integration time, which we believe to be limited by the phase noise of the interferometry laser itself. This value was achieved following the introduction of an FNC setup, which resulted in a factor of five improvement in device performance. Without the FNC system, the gravimeter is limited by the phase noise introduced by the fibre which delivers the laser light to the atoms. These results illustrate the considerable challenges in controlling the interferometry laser phase noise for single-photon gravimeters.


 

For the gravity gradiometer, we experimentally assessed the double lattice launch technique and the scheme for adding arbitrary differential phase for two-cloud atom interferometers. The differential phase imprinted onto the atoms can be obtained by tuning the relative phase between two RF signals that inject into a single AOM on the interferometry laser path. This scheme provides the flexibility to avoid a well-known error arising from use of the ellipse fitting technique to extract small differential phases. Using these methods, we achieve a relative phase sensitivity of 1.53/$\sqrt{\tau}$~rad in the gravity gradiometer configuration.

Our realization and characterization of a single-photon interferometer could open up novel applications for precision measurement and quantum sensing~\cite{safronova2017search}. Combining the advantages of, for example, large-momentum transfer and adiabatic rapid passage with the low power consumption and footprint of semiconductor lasers, would not only relax the complexity requirements of the atom source without precluding the use of such techniques in existing high-performance devices, but would also be of importance for the future development of large-scale atom interferometers based on strontium atoms in space applications. In particular, it could be designed to fill the frequency bandwidth gap between space-borne laser interferometer detectors like LISA~\cite{lisaWhitePaper_2017} and ground-based instruments like LIGO and Virgo~\cite{abbott2016observation, harry2010advanced}.

\section{Acknowledgments} 
We acknowledge financial support from INFN and the Italian Ministry of Education, University and Research (MIUR) under the Progetto Premiale ``Interferometro  Atomico'' and PRIN 2015. This project has received funding from the Initial Training Network (ITN) supported by the European Commission's 7th Framework Programme under Grant Agreement No. 607493. L. H. acknowledges support by Kayser Italia s.r.l.. N. P. acknowledges financial support from European Research Council, Grant No. 772126 (TICTOCGRAV) and from European Union’s Horizon 2020 Programme, under the project TAIOL of QuantERA ERA-NET Cofund in Quantum Technologies (Grant Agreement No. 731473). E. W. acknowledges financial support from the program of China Scholarship Council (No.201703170201). We thank Grzegorz D. Pekala for his assistance in data collection and analysis.

\section*{References}

\bibliography{AI2}
\bibliographystyle{iopart-num}	

\end{document}